\documentclass[a4paper]{article}
\usepackage{amsmath}
\usepackage[english]{babel}
\usepackage{latexsym}
\usepackage{amssymb}
\usepackage{amscd}
\usepackage{amsgen,amstext,amsbsy,amsopn}

\usepackage{graphicx}

\begin{document}
\title{{Generalized Eigenfunctions for Dirac Operators Near Criticality}}

\author{Peter Pickl\footnote{Institut f\"ur theoretische Physik, Universit\"{a}t
        Wien, Boltzmanngasse 5, 1090 Vienna, Austria
         E-mail: pickl@mathematik.uni-muenchen.de}}
\date{\today}

\newtheorem{thm}{Theorem}[section]
\newtheorem{corollary}[thm]{Corollary}
\newtheorem{lemma}[thm]{Lemma}
\newtheorem{proof}[thm]{Proof}
\newtheorem{prop}[thm]{Proposition}
\newtheorem{definition}[thm]{Definition}
\newtheorem{rem}[thm]{Remark}
\newtheorem{notation}[thm]{Notation}

\newcommand{\dk}{\partial_k}
\newcommand{\dt}{\partial_t}

\newcommand{\RM}{\mathbb{R}}
\newcommand{\NM}{\mathbb{N}}
\newcommand{\CM}{\mathbb{C}}

\newcommand{\B}{{\mathcal{B}}}
\newcommand{\M}{{\mathcal{M}}}
\newcommand{\Mbot}{{\mathcal{M^\bot}}}

\newcommand{\N}{\mathcal{M}}

\newcommand{\quer}{\hspace{-0.2cm}/}

\newcommand{\ed}{\bf}
\newcommand{\eed}{\rm}

\maketitle

\begin{abstract}
Critical Dirac operators are those which have eigenfunctions
and/or resonances for $E=m$. We estimate the behavior of the {\em
generalized} eigenfunctions of critical Dirac operators under
small perturbations of the potential. The estimates are done in
the $L^\infty$-norm. We show that for small $k$ the generalized
eigenfunctions are in leading order multiples of the respective
eigenfunctions and/or resonances. We also estimate the
$k$-derivatives which are important for estimating decay. The
method also applies for other differential operators (for example
Schr\"odinger operators).

\end{abstract}
\maketitle
\newpage
\tableofcontents
\newpage

\section{Introduction}

Expansion into generalized eigenfunctions of Schr\"odinger- or Dirac
operators is an important technique in physics to get control on the
time evolution of wave functions. Moreover it was used to establish
completeness in scattering theory (see for example \cite{ikebe}) as
well as to establish the so called Flux Across Surfaces Theorem
which lies at the basis of scattering theory
(\cite{daumer,teufel,panati,pickl,moser}). Of particular interest is
however the case which in scattering theory is normally excluded,
namely when resonances occur and/or there is an eigenvalue on the
edge of the continuous spectrum. We say then that the operator is
critical. Such a critical situation occurs very naturally for the
Dirac operator with a very strong time dependent external potential
which is compactly supported. In this case the famous relativistic
effect of pair creation can happen. This is best be pictured by
considering a time dependent eigenvalue of the time dependent Dirac
operator. For not so strong fields eigenvalues may lie within the
spectral gap $(-mc^2,mc^2)$. When the potential adiabatically
increases (decreases) an eigenvalue increases (decreases) and
eventually touches the positive (negative) edge, i.e. $\pm mc^2$
($\widehat{=}\;\;k=0$) of the continuous spectrum. On the edge the
eigenvalue becomes either a resonance or stays an eigenvalue.
Generically on the upper edge the eigenvalue stays an eigenvalue,
while on the lower edge it typically ceases to be an eigenvalue
(\cite{klaus} and \cite{simon} for the Schr\"odinger case). When the
potential increases further, the eigenvalue disappears.

The question then is what happens to the critical bound state, i.e.
the state corresponding to the eigenvalue on the edge. Does it
scatter? If so pair creation is achieved. This situation has been
extensively studied in the physics literature \cite{beck,rein,
gersh,greiner,prl}. It has also been studied in the mathematical
physics literature \cite{nenciu1,nenciu2,prodan} but the
mathematical proof of pair creation was still lacking until
recently. In \cite{pdneu} we provide a proof of the effect of pair
creation in an adiabatically changing potential where the scattering
behavior of the critical bound state is controlled by generalized
eigenfunctions using the results of the present paper in an
essential way. This then is the physical setting of the problem
studied here. What one needs and what we provide here is the control
of the behavior of the generalized eigenfunctions around and at
criticality. The reader should have in mind the stationary phase
argument to understand what kind of control one needs to control the
time evolution of wave functions. One needs {\em upper bounds} on
the $L^\infty$-norm on the $k-$derivatives of the generalized
eigenfunctions. The quality of that bound is essential for the
quality of the bound on the ``speed of decay'' of the wave function,
see e.g. (\ref{vorschau}) below.

We emphasize that the relevance of this question is by no means
restricted to the Dirac operator case which we just discussed. While
our method can be applied as well to other operators we formulate
our results and proof for the Dirac case because we have the
particular application of pair creation in mind. Moreover we shall
use the Green's function of the free operator in some essential way,
which is more complicated in the Dirac case than in the
Schr\"odinger case: The Green's function of the Dirac Operator is a
matrix-multiple of the Green's function of the Laplacian plus some
extra terms. That is the Schr\"odinger case can be handled following
the Dirac case by essentially omission of some extra terms.

Previous works deal exclusively  with the resolvent \cite{jensen} or
directly with eigenfunctions \cite{tomio} at criticality, but
nothing is known about the behavior in the neighborhood of
criticality, which is generically the relevant question.

From these results it is clear that the normalized (which means
normalized to delta functions in this case) generalized
eigenfunctions of a critical potential diverge as $k$ goes to zero.
We need to generalize this to a family of operators the members of
which vary in the neighborhood of a critical potential. In fact one
should think of the family as arising from the perturbations of a
critical potential. We need to control the behavior of the
generalized eigenfunctions in dependence of $k$ and the perturbation
$B$ of the critical potential. To be clear the generalized
eigenfunctions depend on $k$ and $B$.

The main result of this paper is Theorem \ref{theo}, where we give
an estimate of the $L^\infty$-norm of the normalized generalized
eigenfunctions in dependence of $k$ and $B$ when the critical
potential has a bound state at the edge.

The behavior is different from the case when the critical
potential has no bound state at the edge. The latter situation is
also dealt with in this paper. The result is spelled out in
Theorem \ref{theob}.

Recently in \cite{jennen} a question similar to ours has been asked, namely to estimate the decay of a critical
bound state. While our method is different, it is more general then \cite{jennen} and gives, concerning the
decay, the same result \cite{diss}, \cite{prl}, \cite{pdneu}.

We shall now be more detailed. We shall use units where
$c=m=\hbar=1$. Furthermore throughout the paper the letters $C$
and $C_n$, $n\in\mathbb{N}_0$ will be used for various constants
that need
 not be identical even within the same equation. Finally the absolute value of any vector $\mathbf{x}\in\mathbb{R}^3$
 shall be denoted by $x$.

The one particle Dirac operator $D$ with external potential in the
"standard representation" is defined by
\begin{equation}\label{dirac1}
  D \psi=-i\sum_{l=1}^{3}\alpha_{l}\partial_{l}\psi+A\psi+\beta
  \psi\equiv(D^0+A)\psi\;,
\end{equation}
where the $4\times 4$-matrices $\alpha_l$ and $\beta$ are defined
via
\begin{eqnarray}\label{alphas}
\alpha_{l}=
\begin{pmatrix}
  _{0} & _{\sigma_{l}}\\
  _{\sigma_{l}} & _{0}
\end{pmatrix}; \beta=
\begin{pmatrix}
  _{\mathbf{1}} & _{0}\\
  _{0} & _{-\mathbf{1}}
\end{pmatrix}; l=1,2,3
\end{eqnarray} with $\sigma_{l}$ being the Pauli matrices
\begin{eqnarray*}
 \sigma_{1}=\begin{pmatrix}
  _{1} & _{0} \\
  _{0} & _{-1}
\end{pmatrix}; \sigma_{2}=\begin{pmatrix}
  _{0} & _{1} \\
  _{1} & _{0}
\end{pmatrix}; \sigma_{3}=\begin{pmatrix}
  _{0} & _{-i} \\
  _{i} & _{0}\;
\end{pmatrix}
\end{eqnarray*}
and
\begin{equation}\label{potentialA}
A=A_0 +\sum_{l=1}^{3}\alpha_{l}A_{l}
\end{equation}
for the four potential $A_\mu $ ($A$ is usually denoted by
$A\quer$ in the literature).

Note that $\psi$ is a 4-vector valued function and the underlying
Hilbert space is $\mathcal{H}=L^2(\mathbb{R}^3)^4$.

We are interested in the (generalized) eigenfunctions of the Dirac operator, i.e. $L^\infty$-solutions of
\begin{equation}\label{dirac}
E\phi_E=D\phi_E
\end{equation}
for $E\in\mathbb{R}$.

One can show (see Lemma \ref{boundstates} below), that for a rather
general class of potentials $A$ any such solution solves the so
called Lippmann Schwinger equation
\begin{equation}\label{LSE}
\phi_E(\mathbf{x})=\chi_E(\mathbf{x})+\int
 G^{+}_{E}(\mathbf{x}-\mathbf{y}) A(\mathbf{y})
 \phi_E(\mathbf{y})d^{3}y\;,
 \end{equation}
where $G^{+}_{E}$ are the kernels of
$(E-D^0)^{-1}=\lim_{\delta\rightarrow0}(E-D^0+ i\delta)^{-1}$
and the $\chi_E\in L^\infty$ are solutions of
\begin{equation}\label{freedirac}
E\chi_E=D^0\chi_E\;.
\end{equation}

Let us heuristically explain the main point of this paper. We are
interested in the behavior of the $L^\infty$-norm of the
$L^\infty$-solutions of (\ref{LSE}) with energy
$E_k=\pm\sqrt{k^2+1}$ for critical potential $A$ plus some small
perturbation $B$. The $L^\infty$-solutions of (\ref{freedirac}) for
$E_k=\pm\sqrt{k^2+1}$ are $e^{i\mathbf{k}\cdot\mathbf{x}}$
multiplied with some ($\mathbf{k}$-dependent) spinor. For any
$\mathbf{k}\in\mathbb{R}^3$ and any sign of $E$ there exist two
different $L^\infty$-normalized $\chi(j,\mathbf{k},\cdot)$ (spin
degeneration, see \cite{thaller}). To distinguish between these
different solutions we have introduced the spin index $j$ which is
$1$ or $2$ for positive energies and $3$ or $4$ for negative
energies.

It is already known (see \cite{pickl}) that for any $B$ and any
$\chi(j,\mathbf{k},\cdot)$ (so for any
$(\mathbf{k},j)\in\mathbb{R}^3\times\{1,2,3,4\}$) there exists (up
to linearity) exactly one solution $\phi(A+B,j,\mathbf{k},\cdot)$ of
(\ref{LSE}). We have (see again \cite{pickl}) for non-critical $A+B$
that
$$\sup_{(\mathbf{k},j)\in \mathbb{R}^3\times\{1,2,3,4\}}\|\phi(A+B,j,\mathbf{k},\cdot)\|_\infty<\infty\;,$$
but for $B\equiv0$ (see \cite{jensen})
$$\lim_{k\rightarrow 0}\sup_{j=1,2,3,4}\|\phi(A,j,\mathbf{k},\cdot)\|_\infty=\infty\;.$$
The central part of this paper is to generalize this result and
estimate the $B$ and $k$ behavior of
$\|\phi(A+B,j,\mathbf{k},\cdot)\|_\infty$ for $(B,\mathbf{k})$
around $(0,0)$. We will show, that in the generic case, which means
that the Dirac operator with potential $A$ has a bound state
$\Phi\in L^2$ with energy $1$ or $-1$ (for simplicity we give the
formula in the case that the bound state is non-degenerate)
\begin{equation}\label{vorschau}\|\phi(A+B,j,\mathbf{k},\cdot)\|_\infty\leq\frac{Ck}{\left|\left\langle
\Phi,B,\Phi\right\rangle-C_2k^2\right|+k^3}\end{equation} for some
real constants $C,C_2,C_3$ uniform in $\mathbf{k}$ and $B$ (c.f.
Corollary \ref{supercor}).

This result is an important step forward in controlling the propagation of wave functions under the influence of
critical potentials with small perturbations via eigenfunction expansion. One application of this is the decay
of the QED vacuum via spontaneous (=adiabatic) pair creation under the influence of an adiabatic external
potential. Adiabatic pair creation occurs just when the external potential becomes overcritical, so
(\ref{vorschau}) is useful to estimate the rate and the momentum spectrum of the pairs.

The paper is organized as follows. In the following section 3 we
give the Lippmann Schwinger equation adapted to our setting and give
some relevant formulas. In section \ref{sol} we formulate the main
results as well as a Corollary which is formulated for a somewhat
easier situation and which has to grasp the meaning of the main
result. In section \ref{discuss} we discuss the physical meaning of
the Corollary. Section \ref{givesproof} gives the proof, in section
\ref{sectionkder} we generalize our result to the $k$-derivatives of
the generalized eigenfunctions which are needed to achieve good
control on the time evolution of wave functions (stationary phase).
Some technical details have been put into the appendix.

\section{Solutions of the Lippmann Schwinger Equation}\label{sol}

In view of (\ref{LSE}) we define
\begin{definition}

Let $\mathcal{B}\subset L^\infty$ be the Banach space of functions tending uniformly to zero for
$x\rightarrow\infty$. Let for $A\in L^1\cap L^\infty$ and $E\in\mathbb{R}$ the $T^A_E:L^\infty\rightarrow
\mathcal{B}$ be the operator defined by

\begin{eqnarray}\label{deft}
T^A_E f(\mathbf{x})\nonumber&:=&\int
 G^{+}_{E}(\mathbf{y})A(\mathbf{x}-\mathbf{y})f(\mathbf{x}-\mathbf{y})d^3y
 \\\label{deft2}
&=&-\int
 G^{+}_{E}(\mathbf{x}-\mathbf{y})A(\mathbf{y})f(\mathbf{y})d^3y\;.
\end{eqnarray}
\end{definition}
By this definition (\ref{LSE}) can be written as
\begin{equation}\label{LSE2}
(1- T^A_E)\phi_E(\mathbf{x})=\chi_E(\mathbf{x})\;,
\end{equation}
furthermore
\begin{equation}\label{Tlin}
T_E^{A+B}=T_E^A+T_E^B\;.
\end{equation}
Note that for $|E|<1$ there exists only the trivial solution of the
free Dirac eigenvalue equation (\ref{freedirac}), hence for $|E|<1$
(\ref{LSE2}) reads
\begin{equation}\label{bsolution}
(1- T^A_E)\phi_E=0\;.
\end{equation}
The proof that $T_E^A$ maps $L^\infty$ into $\mathcal{B}$ can be
found in \cite{pickl}.







\begin{lemma}\label{boundstates}
Let $A\in L^\infty\cap L^1$ be H\"older continuous of degree one. Then
\begin{itemize}

\item[(a)] Any $\phi_E\in L^\infty$ satisfies (\ref{dirac}) if and only
if $(1-T_E^A)\phi_E$ satisfies (\ref{freedirac}),

\item[(b)] for any solution $\phi_E\in\mathcal{B}$ of (\ref{LSE2}) we have that $|E|\leq 1$
and thus $\Phi_E$ satisfies (\ref{bsolution}).

\item[(c)] for any nontrivial solution $\phi_E\in L^\infty\backslash\mathcal{B}$ of (\ref{LSE2}) we have that $|E|\geq 1$
and $\Phi_E$ satisfies (\ref{LSE2}) with nontrivial $\chi_E$.


\end{itemize}

\end{lemma}

\noindent\textbf{Proof:} The statement (a) is well know. Since the
proof is short and easy we shall give it now. Let $A\in L^\infty\cap
L^1$ be H\"older continuous of degree one. Note that any $L^\infty$
solution of (\ref{dirac}) is  continuous (even partially
differentiable, since $D^0 f\in L^\infty$). Furthermore $T_E^A f$ is
continuous  for any $f\in L^\infty$ (see for example \cite{pickl}),
hence any solution of (\ref{LSE2}) is continuous.

By definition of $T_E^A$ we have for any continuous $f\in L^\infty$
that
$$(E-D^0)T_E^A f=\int \delta(\mathbf{x}-\mathbf{y})A(\mathbf{y})f(\mathbf{y})d^3y=Af\;,$$ hence
$$(D^0-E)(1-T_E^A) f= (D^0-E+A) f=(D-E)f\;.$$
It follows, that $(D-E)f=0 \Leftrightarrow (D^0-E)(1-T_E^A) f=0$. In
other words, $f$ is solution of (\ref{dirac}) if and only if
$(1-T_E^A) f$ is solution of (\ref{freedirac}), i.e. $f$ solves
(\ref{LSE2}) with some $\chi_E$ solving (\ref{freedirac}).

The proof of (b) is as follows: Let $\phi_E\in\mathcal{B}$ be
solution of (\ref{LSE}). Since $T^A_E$ maps $L^\infty$ into
$\mathcal{B}$ it follows that $\chi_E(\mathbf{x})\in\mathcal{B}$.
Since  there exist no solutions $\chi_E(\mathbf{x})\in\mathcal{B}$
of (\ref{freedirac}) but the trivial one, it follows that
$\chi_E(\mathbf{x})\equiv0$. With (\ref{LSE}) we get
(\ref{bsolution}). Due to \cite{thaller} no solutions of
(\ref{bsolution}) exist for the potentials we consider for $|E|>1$
and (b) follows.

For (c) recall that $T_E^A$ maps $L^\infty\to\\mathcal{B}$, hence
$\phi_E\in L^\infty\backslash\mathcal{B}$ implies that $\chi_E\in
L^\infty\backslash\mathcal{B}$, in particular $\chi_E$ is
nontrivial. Nontrivial solutions of (\ref{freedirac}) exist only for
$|E|\geq 1$ and (c) follows. $\square$

\subsection{Critical Potentials}

In view of Lemma \ref{boundstates} we have that for $|E|>1$ there
only exist solutions of (\ref{dirac}) which also solve (\ref{LSE2})
with nontrivial $\chi_E$. For $|E|<1$ there only exist solutions of
(\ref{dirac}) which also solve (\ref{bsolution}). For $E=\pm 1$ -
depending on the potential $A$ - both kinds of eigenfunctions may
exist. It is known, that generically solutions of (\ref{dirac}) with
$E=\pm 1$ also solve (\ref{LSE2}) with nontrivial $\chi_{\pm 1}$.

Potentials where the $E=\pm 1$ solutions of (\ref{dirac}) also solve
(\ref{bsolution}) are called critical in the literature. We will
focus on positive energy only, so ``critical'' means here, that
there is a $E=+1$ solution of (\ref{dirac}) which solves
(\ref{bsolution}). All results can be obtained equivalently for
negative energies, too (see Remark \ref{vorzeich}).

\begin{definition}\label{critical}
We call a 4-potential $A$ critical if and only if there exist
solutions $\Phi$ of (\ref{boundstates}) with energy $E=1$ (i.e. $(1-
T^A_{1})\Phi=0$). We denote the set of these solutions by
$\mathcal{N}$

\begin{equation}\label{defmengen}
\mathcal{N}:=\{\Phi\in\mathcal{B}:(1- T^A_{1})\Phi=0\}\;.
\end{equation}
\end{definition}

The elements of $\mathcal{N}$ can be bound states (i.e.
$L^2$-solutions of (\ref{bsolution})) or so called resonances (i.e.
not square integrable $\mathcal{B}$-solutions of (\ref{bsolution})).
Next we shall find a formula which distinguishes between these two
different cases and which shall play a crucial role later on.

Let $\Phi\in\mathcal{N}$, i.e.
\begin{eqnarray*}
 \nonumber\Phi(\mathbf{x})&=&\int
 G^{+}_1(\mathbf{y}) A(\mathbf{x}-\mathbf{y})
 \Phi(\mathbf{x}-\mathbf{y})d^{3}y\;.
\end{eqnarray*}
The explicit form of $G^{+}_E$ can be found in \cite{thaller}
\begin{equation}\label{kernel}
G^{+}_{E}(\mathbf{x})=\frac{1}{4\pi}e^{ikx}\left(-x^{-1}(E_{k}+\sum_{j=1}^{3}\alpha_{j}k\frac{x_{j}}{x}+\beta)
-ix^{-2}\sum_{j=1}^{3}\alpha_{j}\frac{x_{j}}{x}\right)\;,
\end{equation}
where $k=\sqrt{E^2-1}$ (hence $E=1$ implies $k=0$). Thus
\begin{eqnarray}\label{vorwirdnull1}
 \Phi(\mathbf{x})
 &=&-\int
\frac{1}{4\pi}y^{-1}(1+\beta+i\sum_{j=1}^{3}\alpha_{j}\frac{y_{j}}{y^2})A(\mathbf{x}-\mathbf{y})
 \Phi(\mathbf{x}-\mathbf{y})d^{3}y
\nonumber\\&=&-\int \frac{1}{4\pi}\left((y^{-1}-x^{-1})(1+\beta)
+i\sum_{j=1}^{3}\alpha_{j}\frac{y_{j}}{y^3})\right)A(\mathbf{x}-\mathbf{y})
 \Phi(\mathbf{x}-\mathbf{y})d^{3}y
\nonumber\\&&-x^{-1}\int \frac{1}{4\pi}(1+\beta )A(\mathbf{x}-\mathbf{y})
 \Phi(\mathbf{x}-\mathbf{y})d^{3}y
\nonumber\\&=:&\Phi_1(\mathbf{x})+\Phi_2(\mathbf{x})\;.
\end{eqnarray}

One can show that for large $x$ the $\Phi_1$ decays at least as fast
as $x^{-2}$. The heuristics for that is rather clear. $A$ decays
fast, thus for large $x$ the integrand is negligible if
$|\mathbf{y}-\mathbf{x}|\gg 1$. Thus the factor
$y^{-1}-x^{-1}=(x-y)/(xy)$ is for large $x$ of order $x^{-2}$, so
$\Phi_1(\mathbf{x})$ decays at least as fast as $x^{-2}$. A rigorous
proof for that is given in the appendix.

To find out, whether $\Phi\in L^2$ it is left to control $\Phi_2(\mathbf{x})$. The decay of $\Phi_2(\mathbf{x})$
depends on the spinor components of $\Phi(\mathbf{y})$. Setting
\begin{equation}\label{wirdnull}\lambda(\Phi):=\int (1+\beta) A(\mathbf{y})
 \Phi(\mathbf{y})d^{3}y\end{equation}
there are two alternatives: Either the spinor components of
$\Phi(\mathbf{y})$ are such that $\lambda(\Phi) \neq 0$ and thus
$\Phi_2(\mathbf{x})$ is of order $x^{-1}$ and thus $\Phi\notin L^2$
or such that $\lambda(\Phi) =0$ and thus $\Phi\in L^2$. The final
result of this paper will depend on whether $\lambda(\Phi) $ is
equal to zero or not, i.e. if $\Phi\in L^2$ or not.

This dichotomy can be compared to the results of \cite{klaus}, where the behavior of bound states of an almost
critical potential is studied. This behavior crucially depends on the fact if $\lambda =0$ or not. Further
explanation how this is related to our results shall be given below.

\begin{notation}
Below we will restrict ourselves to potentials where either
$\lambda(\Phi)=0$ for all $\Phi\in\mathcal{N}$, or
$\lambda(\Phi)\neq0$ for all $\Phi\in\mathcal{N}$. To distinguish
between these two cases we define
$$\overline{\lambda}:=\left\{
                      \begin{array}{ll}
                        0, & \hbox{if $\lambda(\Phi)=0$
for all $\Phi\in\mathcal{N}$;} \\
                        1, & \hbox{is $\lambda(\Phi)\neq0$ for all $\Phi\in\mathcal{N}$.}
                      \end{array}
                    \right.$$
\end{notation}
This restriction rules out potentials with  $\dim \mathcal{N}>2$ and
$\lambda(\Phi)\neq0$: If $\dim \mathcal{N}>2$ one can always find a
$\Phi\in\mathcal{N}$ such that $\lambda(\Phi)=0$ using linearity of
(\ref{wirdnull}) and the fact that the kernel of $1+\beta$ is two
dimensional (hence the image of $1+\beta$ is two dimensional so
$\lambda(\Phi)$ is always an element of a two dimensional subspace
of $\mathbb{C}^4$).


\section{Generalized Eigenfunctions for Critical Potentials with
Small Perturbations}

\begin{definition}\label{niceB}

For any selfadjoint matrix valued multiplication operator $A\in L^1$
let the (pseudo) scalar product $\left\langle
\cdot,A,\cdot\right\rangle:L^\infty\times
L^\infty\rightarrow\mathbb{C}$ be given by
$$\left\langle
f,A,g\right\rangle:=\int
f^\dagger(\mathbf{x})A(\mathbf{x})g(\mathbf{x})d^3x\;.$$ For any
$K>0$ let the set $\mathcal{W}_K\subset L^\infty\cap L^1$ be given
by
$$B\in\mathcal{W}_K\Leftrightarrow B\in L^\infty\cap L^1\text{ with }\frac{\|\Phi\|_\infty^2(\| B\|_{1}+\|B\|_\infty)^2}{|\left\langle \Phi, B, \Phi \right\rangle|}\leq K\text{ for all }\Phi\in\mathcal{N}\;.$$
For any critical $A\in L^1\cap L^\infty$ we define the following
subspaces of $\mathcal{B}$
\begin{eqnarray*}
\mathcal{M}^\parallel&:=&A\mathcal{N}:=\{A\Phi:\Phi\in\mathcal{N}\}\subset
L^2\\ \mathcal{M}^\bot&:=&\{m^\bot\in\mathcal{B}:\left\langle
m^\bot,A,\Phi\right\rangle=0 \;\forall\; \Phi\in\mathcal{N}\}\;.
\end{eqnarray*}
\end{definition}
In the following we will restrict our observations to critical potentials which satisfy some additional (weak)
conditions.
\begin{definition}\label{nicecritical}
Let $\mathcal{C}$ be the set of critical potentials defined by
$A\in\mathcal{C}$ if and only if
\begin{description}
    \item[(a)] $A$ is critical and H\"older continuous of degree one,
    \item[(b)] $(1+|\cdot|)^2A\in L^1\cap L^\infty$,
    \item[(c)] $\mathcal{N}\cap \mathcal{M}^\bot=\{0\}$,
    \item[(d)] either $\mathcal{N}\subset L^2$ or $\mathcal{N}\cap L^2=\{0\}$,
    \item[(e)] either
\begin{equation}\label{restricta1}
\int A(\mathbf{x})\Phi(\mathbf{x})d^3x\neq0
\end{equation}
or
\begin{equation}\label{restricta2}
(1-i\beta)\int A(\mathbf{x})\Phi(\mathbf{x})\mathbf{x}d^3x\neq0\;.
\end{equation}
\end{description}
\end{definition}
\begin{rem}

It is rather clear that either (\ref{restricta1}) or (\ref{restricta2}) are satisfied for almost every critical
potential. For example if $\Phi$ is a ground state and $A$ is purely electric (=multiple of the unit matrix) and
positive, the Perron–-Frobenius Theorem implies that (\ref{restricta1}) holds.

Furthermore we have for any purely electric, positive critical,
``short range'' potential (which means in our case
$(1+|\cdot|)^2A\in L^1\cap L^\infty$) that $\mathcal{N}\cap
\mathcal{M}^\bot=\{0\}$: Obviously $\left\langle\Phi, A,
\Phi\right\rangle>0$ for any positive electric
 potential $A$ and any
$\Phi\in\mathcal{N}$. Small perturbations do not significantly
change $\left\langle\Phi, A, \Phi\right\rangle$. Hence the set of
critical, ``short range'' potentials with $\mathcal{N}\cap
\mathcal{M}^\bot=\{0\}$ is not small. It seems that almost every
critical ``short range'' (in the given sense) potential lies in
$\mathcal{C}$.
\end{rem}
In this paper we wish to estimate the generalized eigenfunctions of
the Dirac operator with potentials $A+B$ where $A\in\mathcal{C}$ and
$B\in \mathcal{W}_K$ for some (small) $K$. The generalized
eigenfunctions are the respective solutions of (\ref{LSE2}), i.e.
solutions of
\begin{equation}\label{LSE3}
(1- T^{A+B}_{\pm
E_k})\phi(A+B,j,\mathbf{k},\cdot)=\chi(j,\mathbf{k},\cdot)\;,
\end{equation}
where the $\chi(j,\mathbf{k},\cdot)$ are the $L^\infty$-normalized
generalized eigenfunctions of the free Dirac operator with momentum
$\mathbf{k}$ and spin $j$, $E_k=\sqrt{k^2+1}$ and the sign $+$ holds
for $j=1,2$, the sign $-$ holds for $j=3,4$. For ``small'' $B$ we
have - similar as in the $B=0$-case (see \cite{jensen}) that the
generalized eigenfunctions are of leading order a multiple of some
element of $\mathcal{N}$. Which element may depend on $B,\mathbf{k}$
and $j$. We will estimate the divergent behavior of this element in
dependence of $B,\mathbf{k}$ and $j$ and the $L^\infty$-norm of the
generalized eigenfunctions minus their leading order
$\mathcal{N}$-part. As mentioned above, that behavior of the
generalized eigenfunctions depends crucially on the fact is
$\overline{\lambda}=0$ or $\overline{\lambda}=1$. It is convenient
to give two Theorems separating these two different cases. For
$\overline{\lambda}=0$ we have
\begin{thm}\label{theo} 
Let  $A\in\mathcal{C}$ with $\overline{\lambda}=0$ (i.e.
$\mathcal{N}\subset L^2$). Then there exist constants
 $C,K,k_0>0$ and a selfadjoint linear map $\widehat{R}:\mathcal{N}\to\mathcal{N}$
 such that for any
$\mathbf{k}\in\mathbb{R}^3$ with $k<k_0$, $j=1,2$, any potential
$B\in\mathcal{W}_K$ there exists a
$\Phi^B_{j,\mathbf{k}}\in\mathcal{N}$ with
\begin{equation}\label{resultat}\|\Phi^B_{j,\mathbf{k}}\|\leq
C\left(k+\|B\|_1\right)\left(\inf_{\Phi\in
\mathcal{N},\|\Phi\|=1}\left\| \left(P_{\mathcal{N}}^\parallel B+
\widehat{R}k^2\right)\Phi\right\|+k^3
 \right)^{-1}\end{equation}
and (c.f. (\ref{LSE3}))
\begin{equation}\label{resultat2}
\|\phi(A+B,j,\mathbf{k},\cdot)-\Phi^B_{j,\mathbf{k}}\|_\infty<C+Ck\left(\|B\|_1+\|B\|_\infty\right)\left(\inf_{\Phi\in
\mathcal{N},\|\Phi\|=1}\left\| \left(P_{\mathcal{N}}^\parallel B+
\widehat{R}k^2\right)\Phi\right\|+k^3
 \right)^{-1}\;.
\end{equation}
\end{thm}
For $\overline{\lambda}=1$ we have
\begin{thm}\label{theob} 
Let $A\in\mathcal{C}$ and $\overline{\lambda}=1$ (i.e.
$\mathcal{N}\cap L^2=\emptyset$). Then there exist constants
 $C,K,k_0>0$ such that for any
$\mathbf{k}\in\mathbb{R}^3$ with $k<k_0$, $j=1,2$ and any potential
$B\in\mathcal{W}_K$ there exists a
$\Phi^B_{j,\mathbf{k}}\in\mathcal{N}$ with
\begin{eqnarray}\label{resultatb}
\|\Phi^B_{j,\mathbf{k}}\|\leq C\left(\inf_{\Phi\in
\mathcal{N},\|\Phi\|=1}\left|\left\langle
\Phi,B,\Phi\right\rangle\right|+|k|
 \right)^{-1}
\end{eqnarray}
and
\begin{equation}\label{resultat2b}
\|\phi(A+B,j,\mathbf{k},\cdot)-\Phi^B_{j,\mathbf{k}}\|_\infty<C\left(\|B\|_1+\|B\|_\infty\right)\left(\inf_{\Phi\in
\mathcal{N},\|\Phi\|=1}\left|\left\langle
\Phi,B,\Phi\right\rangle\right|+|k|
 \right)^{-1}\;.
\end{equation}
\end{thm}

\begin{rem}\label{vorzeich}
Note, that for any $\mathbf{k}\in\mathbb{R}^3$ there exist two
linearly independent generalized eigenfunctions
$\chi(j,\mathbf{k},\cdot)$ and two linearly independent generalized
eigenfunctions $\chi_{-E_k}$ of the free Dirac operator with energy
$\pm E_k=\pm\sqrt{1+k^2}$. Using CPT-symmetry the Theorem is also
valid for potentials $A$ which are ``critical'' in the sense that
they have bound states or a resonance with energy $-1$. It then
gives estimates on the generalized eigenfunctions with negative
energy (i.e. $j=3,4$) of course.
\end{rem}

To make it easier to understand the statement of Theorem \ref{theo},
let us restrict ourselves on potentials $B_\mu$ which can be written
as $B_\mu=\mu B_0$ for some fixed potential $B_0$ and
$\mu\in[-\mu_0,\mu_0]$. $B_0$ and $\mu_0\in\mathbb{R}^+$ are chosen
such, that $B_\mu\in\mathcal{W}_K$ for all $\mu\in[-\mu_0,\mu_0]$.
Under these restrictions we get
\begin{corollary}\label{supercor}
Let $A\in\mathcal{C}$ with $\overline{\lambda}=0$. Let $B_0\in
L^\infty\cap L^1$ with $\left\langle\Phi, B_0,\Phi\right\rangle\neq
0$ for all $\Phi\in\mathcal{N}\backslash\{0\}$. Then there exist
constants
 $C,\mu_0,k_0>0$ and constants $\gamma_l$, $l=1,\ldots,n=\dim\mathcal{N}$
 such that for any
$\mathbf{k}\in\mathbb{R}^3$ with $k<k_0$, $j=1,2$, any
$\mu\in[-\mu_0,\mu_0]$ there exists a
$\Phi^\mu_{j,\mathbf{k}}\in\mathcal{N}$ with
\begin{equation}\label{resultatc}\|\Phi^\mu_{j,\mathbf{k}}\|\leq
C+Ck\sum_{l=1}^n \left(|\mu+\gamma_lk^2|+k^3
 \right)^{-1}\end{equation}
and
\begin{equation}\label{resultat2c}
\|\phi(A+\mu
B_0,j,\mathbf{k},\cdot)-\Phi^\mu_{j,\mathbf{k}}\|_\infty<C\;.
\end{equation}
\end{corollary}
\noindent\textbf{Proof:} We choose $\mu_0$ such that $\mu
B_0\in\mathcal{W}_K$ for all $\mu\in[-\mu_0,\mu_0]$. Hence the
assumptions of Theorem \ref{theo} hold for $\mu_0$ small enough and
we only need to show that the right hand sides of (\ref{resultat})
and (\ref{resultat2}) are bounded by the right hand sides of
(\ref{resultatc}) and (\ref{resultat2c}) respectively.

For that note that under the given assumptions
\begin{equation}\label{super1}\|\mu B_0\|_1\leq C k+2\|\mu B_0\|_1\left(\inf_{\Psi\in
\mathcal{N},\|\Psi\|=1}\left\| \left(P_{\mathcal{N}}^\parallel \mu
B_0+ \widehat{R}k^2\right)\Psi\right\|+k^3\right)\left(\inf_{\Psi\in
\mathcal{N},\|\Psi\|=1}\left|\left\langle\Psi, \mu
B_0,\Psi\right\rangle\right|\right)^{-1}\end{equation}
 for appropriate
$C<\infty$. This one can prove by considering two different cases.
First assume that $\inf_{\Psi\in\mathcal{N},\|\Psi\|=1}|\left\langle
\Phi, \mu B_0, \Phi\right\rangle|>2\|\widehat{R}\|_{op}k^2 $. It
follows that
\begin{eqnarray*}\left(\inf_{\Psi\in \mathcal{N},\|\Psi\|=1}\left\|
\left(P_{\mathcal{N}}^\parallel \mu
B_0\right)\Psi\right\|\right)+k^3
\leq\left(\inf_{\Psi\in \mathcal{N},\|\Psi\|=1}\left|\left\langle
\Psi,\mu B_0,\Psi\right\rangle\right|\right)
\leq\frac{1}{2}\inf_{\Psi\in\mathcal{N},\|\Psi\|=1}|\left\langle
\Psi, \mu B_0, \Psi\right\rangle|
\end{eqnarray*}
and the second summand of (\ref{super1}) gives an appropriate bound.
Assuming that $\inf_{\Psi\in\mathcal{N},\|\Psi\|=1}|\left\langle
\Phi, \mu B_0, \Phi\right\rangle|<2\|\widehat{R}\|_{op}k^2 $ we have
for $\mu B_0\in\mathcal{W}_K$ (c.f. Definition \ref{niceB}) that
$\|\mu B_0\|_1^2< 2K \|\widehat{R}\|_{op}k^2$ and thus the first
summand of (\ref{super1}) gives an appropriate bound.

Similarly one gets that
\begin{eqnarray}\label{super2}
&&Ck\left(\|\mu B_0\|_1+\|\mu B_0\|_\infty\right)\left(\inf_{\Psi\in
\mathcal{N},\|\Psi\|=1}\left\| \left(P_{\mathcal{N}}^\parallel \mu
B_0+ \widehat{R}k^2\right)\Psi\right\|+k^3
 \right)^{-1}\\\nonumber&&<C\left(\|\mu B_0\|_1+\|\mu B_0\|_\infty\right)\left(\inf_{\Psi\in
\mathcal{N},\|\Psi\|=1}\left|\left\langle\Psi, \mu
B_0,\Psi\right\rangle\right|\right)^{-1}\;.
\end{eqnarray}
Assuming that $\inf_{\Psi\in\mathcal{N},\|\Psi\|=1}|\left\langle
\Phi, \mu B_0, \Phi\right\rangle|>2\|\widehat{R}\|_{op}k^2 $ the
formula can be proven as (\ref{super1}) above, assuming that
$\inf_{\Psi\in\mathcal{N},\|\Psi\|=1}|\left\langle \Phi, \mu B_0,
\Phi\right\rangle|<2\|\widehat{R}\|_{op}k^2 $ we have
\begin{eqnarray*}
&&Ck\left(\|\mu B_0\|_1+\|\mu B_0\|_\infty\right)\left(\inf_{\Psi\in
\mathcal{N},\|\Psi\|=1}\left\| \left(P_{\mathcal{N}}^\parallel \mu
B_0+ \widehat{R}k^2\right)\Psi\right\|+k^3
 \right)^{-1}<C\left(\|\mu B_0\|_1+\|\mu B_0\|_\infty\right)k^{-3}\\&&=C\left(\|\mu B_0\|_1+\|\mu B_0\|_\infty\right)k^{-2}
 \leq C\left(\|\mu B_0\|_1+\|\mu B_0\|_\infty\right)\left(\inf_{\Psi\in
\mathcal{N},\|\Psi\|=1}\left|\left\langle\Psi, \mu
B_0,\Psi\right\rangle\right|\right)^{-1}\leq C\;,
\end{eqnarray*}where we used in the last step that $\left(\|  B_0\|_1+\|  B_0\|_\infty\right)\left(\inf_{\Psi\in
\mathcal{N},\|\Psi\|=1}\left|\left\langle\Psi,
B_0,\Psi\right\rangle\right|\right)^{-1}$ exists by assumptions on
$B_0$. With that and (\ref{resultat2}) equation (\ref{resultat2c})
follows.

Next we prove (\ref{resultatc}). Using (\ref{super1}) in
(\ref{resultat}), noting that $$\|\mu B_0\|_1\left(\inf_{\Psi\in
\mathcal{N},\|\Psi\|=1}\left|\left\langle\Psi, \mu
B_0,\Psi\right\rangle\right|\right)^{-1}=\|
B_0\|_1\left(\inf_{\Psi\in
\mathcal{N},\|\Psi\|=1}\left|\left\langle\Psi,
B_0,\Psi\right\rangle\right|\right)^{-1}<C$$ we have
\begin{equation}\label{resultatz}\|\Phi^\mu_{j,\mathbf{k}}\|\leq C +Ck\left(\inf_{\Phi\in \mathcal{N},\|\Phi\|=1}\left\|
\left(P_{\mathcal{N}}^\parallel \mu B_0+
\widehat{R}k^2\right)\Phi\right\|+k^3
 \right)^{-1}\;.\end{equation}

Recall that $\left\langle\Phi, B_0,\Phi\right\rangle\neq 0$ for all
$\Phi\in\mathcal{N}\backslash\{0\}$, hence in particular $B\Phi\neq
0$ for all $\Phi\in\mathcal{N}\backslash\{0\}$. Thus the matrix
$\widehat{B}_0: \mathcal{N}\to\mathcal{N}$ defined by
$$\widehat{B}_0\Phi=P_\mathcal{N} B\Phi$$
is invertible. Hence we get for (\ref{resultatz})
\begin{eqnarray*}
\|\Phi^B_{j,\mathbf{k}}\|&\leq&
C+Ck\left(\inf_{\Phi\in
\mathcal{N},\|\Phi\|=1}\left\|\widehat{B}_0\left(\mu
+\widehat{B}_0^{-1} \widehat{R}k^2\right)\Phi\right\|+k^3
 \right)^{-1}
\\&\leq&C+Ck\|\widehat{B}_0^{-1}\|_{op}\left(\inf_{\Phi\in
\mathcal{N},\|\Phi\|=1}\left\|\left(\mu +\widehat{B}_0^{-1}
\widehat{R}k^2\right)\Phi\right\|+k^3
 \right)^{-1}\;.
\end{eqnarray*}
Using that $\|\widehat{B}_0^{-1}\|_{op}<\infty$ and defining the
symmetric operator $\widehat{M}:\mathcal{N}\to\mathcal{N}$ and the
antisymmetric operator $\widehat{N}:\mathcal{N}\to\mathcal{N}$ by
$$\widehat{M}:=\frac{1}{2}\left(\widehat{B}_0^{-1} \widehat{R}+\widehat{R}\widehat{B}_0^{-1}\right) $$
and
$$\widehat{N}:=\frac{1}{2}\left(\widehat{B}_0^{-1} \widehat{R}-\widehat{R}\widehat{B}_0^{-1}\right) $$
one gets
\begin{eqnarray*}
\|\Phi^B_{j,\mathbf{k}}\|&\leq&C+Ck\left(\inf_{\Phi\in
\mathcal{N},\|\Phi\|=1}\left\|\left(\mu
+\widehat{M}k^2+\widehat{N}k^2\right)\Phi\right\|+k^3
 \right)^{-1}
\\&\leq&C+Ck\left(\inf_{\Phi\in
\mathcal{N},\|\Phi\|=1}\left|\left\langle \Phi,\left(\mu
+\widehat{M}k^2+\widehat{N}k^2\right)\Phi\right\rangle\right|+k^3
 \right)^{-1}
 \;.
\end{eqnarray*}
Note that for symmetric $\widehat{M}$ the $\left\langle
\Phi,\left(\mu +\widehat{M}k^2\right)\Phi\right\rangle$ is real,
whereas for antisymmetric $\widehat{N}$ the $\left\langle
\Phi,\widehat{N}k^2\Phi\right\rangle$ is imaginary, hence
\begin{eqnarray*}
\|\Phi^B_{j,\mathbf{k}}\|&\leq&C+Ck\left(\inf_{\Phi\in
\mathcal{N},\|\Phi\|=1}\left|\left\langle \Phi,\left(\mu
+\widehat{M}k^2\right)\Phi\right\rangle\right|+k^3
 \right)^{-1}
 \;.
\end{eqnarray*}
Let now $\{\Phi_l:l=1,\ldots,n=\dim \mathcal{N}\}$ be an orthonormal
eigenbasis of $\mathcal{M}$, let $\gamma_l:l=1,\ldots,n$ be the
respective eigenvalues. Note that the minimum of $\left|\left\langle
\Phi,\left(\mu +\widehat{M}k^2\right)\Phi\right\rangle\right|$ is
always realized for an eigenstate of $\mu +\widehat{M}k^2$, thus an
element of $\{\Phi_l\}$. Which element will in general depend on $k$
and $\mu$, thus we have
\begin{eqnarray*}
\|\Phi^B_{j,\mathbf{k}}\|&\leq&C+Ck\left(\inf_{\Phi_l}\left|\left\langle
\Phi_l,\left(\mu
+\widehat{M}k^2\right)\Phi_l\right\rangle\right|+k^3
 \right)^{-1}
\\&\leq&C+Ck\sum_{l=1}^n \left(|\mu+\gamma_lk^2|+k^3
 \right)^{-1}
 \;.
\end{eqnarray*}

$\square$

\section{Discussion of the Result}\label{discuss}

Before proving the Theorem let us shortly clarify the physical
meaning of Corollary \ref{supercor} on a heuristic level.

\begin{itemize}

 \item[(a)] If $\overline{\lambda} =0$ it may happen that the nominator in the right hand side of (\ref{resultatc}) is of order
$k^3$ (namely if $\mu+\gamma_lk^2=0$ for some $1\leq l\leq n$). The
respective $k$'s where this happens
 are usually called ``resonances of the potential $A+B$'' in the physics literature.
 Around the resonance the generalized
eigenfunctions are of order $k^{-2}$.

\begin{figure}[h]\label{bild2}
\begin{center}
\includegraphics*[width=240pt,height=160pt]{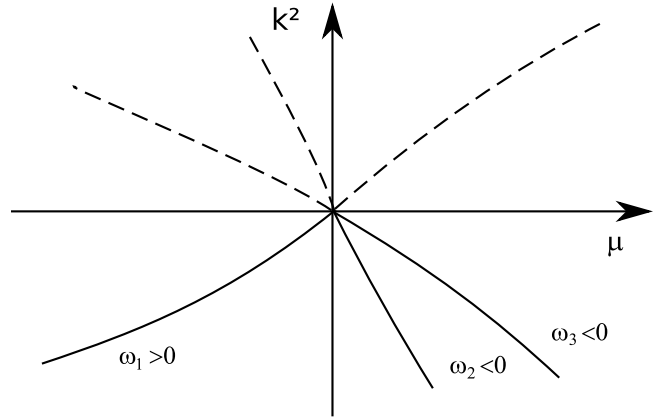} \caption[Bound]{{\bf Bound states and Resonances (Illustration of (b))} The figure illustrates the position of
bound states (illustrated by lines, $k$ is imaginary hence $k^2$
negative) and resonances (dashed lines, $k$ is real hence $k^2$
positive) of the Dirac operator $D=D^0+A+\mu B_0$, with $A$, $B_0$
and $\mu$ as in Corollary \ref{supercor}.}
\end{center}
\end{figure}

\item[(b)] The results of Corollary \ref{supercor} can also be used to roughly  estimate the energy of bound states of
``almost-critical'' potentials $A+B$. Due to Lemma \ref{boundstates}
bound states have energies smaller than $1$, so the respective
$k=i\kappa$ is imaginary. Instead of (\ref{LSE2}) they satisfy
(\ref{bsolution}), implying that $\|\phi_{E_k}\|_\infty=\infty$ for
the respective $B$ and (imaginary) $k$ as one can see as follows.
Heuristically speaking: ``Normalize'' (\ref{LSE2}), i.e. divide
(\ref{LSE2}) on both sides by $\|\phi_{E_k}\|_\infty$. It follows
that (\ref{LSE2}) and (\ref{bsolution}) are equivalent if (and only
if) the right hand side of (\ref{LSE2}) divided by
$\|\phi_{E_k}\|_\infty$ is equal to zero, hence if
$\|\phi_{E_k}\|_\infty=\infty$.

Hence the energy $E_k$ of the $l^{th}$ bound state of the potential
$A+B$ satisfies $\mu+\gamma_lk^2\approx0$ if
 $\overline{\lambda} =0$ (see solid lines in figure \ref{bild2}). This implies, that bound states occur only if the respective $\mu$ and $\gamma_l$ have different sign.
  They ``live'' on different lines with slope $\gamma_l$  through the origin in the $k^2$($\approx E_k-1$) against $\mu$-plot
 (see figure \ref{bild2}).

This estimation is in line with the results of Theorem 1.1 by Klaus (in Klaus' Paper $a$ plays the role of
 $\lambda $) concerning the behavior of the bound state energies at the threshold: $a=0\Leftrightarrow \sigma:=\mu\sim \kappa^2\Leftrightarrow E=\kappa^2$ is not analytic in $\sigma$ (since the next term in the power series is of order $\kappa^3\propto\sigma^{\frac{3}{2}}$ which destroys analyticity);
 $a\neq0\Leftrightarrow \sigma:=\mu\sim \kappa\Leftrightarrow E=\kappa^2$ is analytic in $\sigma$.

This idea is also helpful to find out the sign of the respective
$\gamma_l$: If $B_0$ is such, that there exist (don't exist) bound
states with energy $E_\kappa$ for positive $\mu$ with
$\mu-\gamma_l\kappa^2\approx0$, the respective $\gamma_l$ is
positive (negative) (see again figure \ref{bild2}).

There is physics in this: The fact, that there are bound states
``living'' on different lines comes from the fact, that adding the
potential $B_0$ may destroy the degeneracy of $A$ (For example, if
$A$ was purely electric, thus (at least) spin-degenerated, adding a
small vector potential $B_0$ will in general destroy spin
degeneracy). The degeneracy of the new bound states on each of these
``lines'' is equal to the multiplicity of the respective $\gamma_l$.

It follows, that also the ``resonances'' loose - at least partially
- their degeneracy when a general potential $B_0$ is added. The
estimates (concerning the sum) in Corollary \ref{supercor} reflect
this fact: Each summand represents a ``resonance''. In this sense
one can heuristically guess that the generalized eigenfunction is of
leading order equal to
$$\phi(A+B,j,\mathbf{k},\cdot)\approx\sum_{p=1}^n\frac{Ck}{\mu+\gamma_lk^2
 +iC_3k^3}\Phi_p\;,$$
where the set $\{\Phi_p:1\leq p\leq1\}$ is a basis of
$\mathcal{N}$.

\begin{figure}[h]\label{bild}
\begin{center}
\includegraphics*[width=240pt,height=160pt]{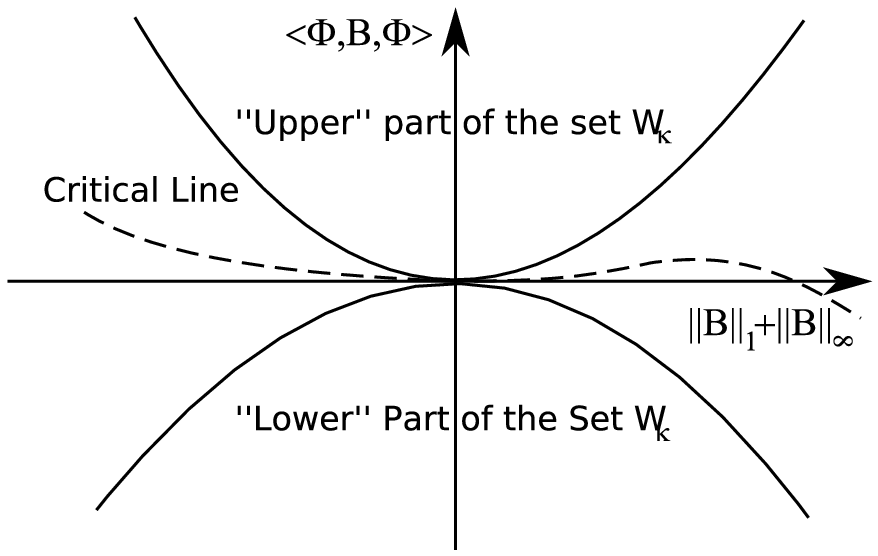} \caption[MengeB1]{{\bf Illustration of the Set $\mathcal{W}_K$
for $n=1$ and normalized $\Phi$}\\

$\mathcal{W}_K$ lies inside two parabolas and is the region where
the Theorem applies. On the ``Critical Line'' are potentials with
critical $A+B$ (see (c)). Both parabolas and the ``critical line''
touch in the origin, implying heuristically that
$\left\langle\Phi,B,\Phi\right\rangle=O((\|B\|_1+\|B\|_\infty)^2)$
for critical $A+B$.

Due to part (b) one of these two parabolas contains potentials which
have bound states, the other one contains potentials which have
``resonances'' (which one, depends on $A$. For positive, purely
electric $A$ on can show that the potentials in the lower
``parabola'' have bound states ).}
\end{center}
\end{figure}

\item[(c)] Let us explain the meaning of the
set $\mathcal{W}_K$ (see figure \ref{bild}). 

Let $A\in\mathcal{C}$. Disturbing $A$ by a small short range potential $B$ it may happen, that $A+B$ stays
critical.

If $A+B$ stays critical, the result of Jensen and Kato gives us,
that the respective generalized eigenfunctions diverge for $k=0$.
Looking at (\ref{resultat}) it follows that in this case either
$\left\langle \Phi, B,
 \Phi
 \right\rangle=0$  for some $\Phi\in\mathcal{N}$ or that the requirements of Theorem
 \ref{theo} are not satisfied, which means that
 $B\notin\mathcal{W}_K$. This fact is a strong requirement on $B$ for the criticality of $A+B$
 (see figure \ref{bild}) for the
non-degenerate case. 
Remember that by definition \ref{niceB}
$B\in\mathcal{W}_K\Leftrightarrow \frac{(\|
B\|_{1}+\|B\|_\infty)^2}{|\left\langle \Phi, B, \Phi
\right\rangle|}\leq K\text{ for all normalized
}\Phi\in\mathcal{N}\Leftrightarrow|\left\langle \Phi, B, \Phi
\right\rangle|\geq K^{-1}(\| B\|_{1}+\|B\|_\infty)^2$. So in the
$\left\langle\Phi,B,\Phi\right\rangle$ against $\|
B\|_1+\|\B|_\infty$ plot, $\mathcal{W}_K$ lies inside two parabolas
with curvature $\pm K$ (see figure \ref{bild}).

\end{itemize}

\section{Proof of the Theorem}\label{givesproof}

The set $\mathcal{M}^\bot$ has the interesting property that it is invariant under $T_1^A$, a fact which will
play a crucial role in what follows

\begin{lemma}

For any $A\in\mathcal{C}$ we have that
$$
h^\bot\in\mathcal{M}^\bot\Leftrightarrow T_1^{A} h^\bot\in\mathcal{M}^\bot\;.
$$
\end{lemma}

\noindent\textbf{Proof:} We show first that for $h,g\in\mathcal{B}$
and $A,B\in L_1$
\begin{equation}\label{symm}
\left\langle h,A,T_E^{B} g\right\rangle = \left\langle T_E^{A} h,
B,g\right\rangle
\end{equation}
by computing
\begin{eqnarray*}
\left\langle h,A, T_E^{B}g \right\rangle&=&\int
h^\dagger(\mathbf{x}) A(\mathbf{x})T_E^{B}g(\mathbf{x})d^3x
\nonumber\\&=&\int h^\dagger(\mathbf{x})A(\mathbf{x})\int
G^{+}_E(\mathbf{x}-\mathbf{y}) B(\mathbf{y})g(\mathbf{y})d^3yd^3x
\nonumber\\&=&\int\int h^\dagger(\mathbf{x})A(\mathbf{x})
G^{+}_E(\mathbf{x}-\mathbf{y})d^3x B(\mathbf{y})g(\mathbf{y})d^3y
\nonumber
\\&=&\int (T^A_{E_k}h)^\dagger(\mathbf{y})B(\mathbf{y})g(\mathbf{y})d^3y \nonumber
\\&=&\left\langle T_E^{A} h, B,g\right\rangle\;.
\end{eqnarray*}
We may apply this to $h\in\mathcal{B}$ and $g=\Phi\in\mathcal{N}$
to obtain
\begin{eqnarray*}
\left\langle h,A,\Phi\right\rangle=\left\langle
h,A,T_1^{A}\Phi\right\rangle=\left\langle
T_1^{A}h,A,\Phi\right\rangle\;.
\end{eqnarray*}
This equation directly implies the Lemma: If $h\in\mathcal{M}^\bot$
(which means that $\left\langle h,A,\Phi\right\rangle=0$) it follows
that $T_1^{A} h\in \mathcal{M}^\bot$ (which means $\left\langle
T_1^{A} h,A, \Phi\right\rangle=0$) and vice versus.

$\square$

Furthermore we have

\begin{lemma}\label{directsum}Using the definitions above we have
\begin{itemize}\item[(a)]
\begin{equation}\label{bissum}
\mathcal{B}=\mathcal{M}^\parallel\oplus\mathcal{M}^\bot\;,
\end{equation}
\item[(b)]
\begin{equation}\label{bissum2}
\mathcal{B}=\mathcal{N}\oplus\mathcal{M}^\bot\;.
\end{equation}
\end{itemize}
\end{lemma}

\begin{rem}
Note that $\left\langle A\Phi,A,\Phi\right\rangle>0$, hence
$\mathcal{M}^\parallel\cap\mathcal{M}^\bot=\{0\}$.

Using that $\mathcal{M}^\parallel\cap\mathcal{M}^\bot=\{0\}$, part (a) of the Lemma defines projectors
$P^\parallel_\mathcal{M}$ and $P^\bot_\mathcal{M}$ with
$P^\parallel_\mathcal{M}\mathcal{B}\subset\mathcal{M}^\parallel$,
$P^\bot_\mathcal{M}\mathcal{B}\subset\mathcal{M}^\bot$ and $P^\parallel_\mathcal{M}+P^\bot_\mathcal{M}=1$.

Using that $\mathcal{N}\cap\mathcal{M}^\bot=\{0\}$ (see Definition \ref{nicecritical}), (b) defines projectors
$P^\parallel_\mathcal{N}$ and $P^\bot_\mathcal{N}$ with $P^\parallel_\mathcal{N}\mathcal{B}\subset\mathcal{N}$,
$P^\bot_\mathcal{N}\mathcal{B}\subset\mathcal{M}^\bot$ and $P^\parallel_\mathcal{N}+P^\bot_\mathcal{N}=1$.
\end{rem}

\begin{description}
\item[Proof (a), (b) ``$\supset$'']
Since $\mathcal{M}^\parallel, \mathcal{M}^\bot,\mathcal{N}\subset
\mathcal{B}$ it follows that
$\mathcal{B}\supset\mathcal{M}^\parallel\oplus\mathcal{M}^\bot$ and
$\mathcal{B}\supset\mathcal{N}^\parallel\oplus\mathcal{M}^\bot$.

\item[(a) ``$\subset$''] Let $f\in\mathcal{B}$, $\{\Phi_p\}$ $p=1\ldots
n$ be a basis of $\mathcal{N}$. Define the vector
$\overrightarrow{f}\in\mathbb{R}^n$ by $f_p:=\left\langle
f,A,\Phi_p\right\rangle$ and for any $q=1\ldots n$ the vector
$\overrightarrow{\Phi}^q$ by $\Phi^q_p:=\left\langle A\Phi_q,
A,\Phi_p\right\rangle$.

We will show by contradiction that the $\overrightarrow{\Phi}^q$ are
linearly independent. Assume that the vectors
$\overrightarrow{\Phi}^q$ are linearly dependent, i.e. that it is
possible to find non-trivial complex numbers $\gamma_q$, $q=1\ldots
n$ such that $0=\sum_{q=1}^n \gamma_j\overrightarrow{\Phi}^q$. In
other words $\left\langle A\sum_{q=1}^n \gamma_q\Phi_q,
A,\Phi_p\right\rangle=0$ for all $p=1\ldots n$, hence $A\sum_{q=1}^n
\gamma_q\Phi_q\in\mathcal{M}^\bot$. Furthermore we have by
definition of $\mathcal{M}^\parallel$ that $A\sum_{q=1}^n
\gamma_q\Phi_q\in\mathcal{M}^\parallel$. Hence
$\mathcal{M}^\parallel\cap\mathcal{M}^\bot=\{0\}$ implies
$A\sum_{q=1}^n \gamma_q\Phi_q\equiv0$ (and thus
$\sum_{q=1}^n\gamma_q\Phi_q=0$ on the support of $A$). But the only
eigenfunction which is equal to zero on the support of $A$ is
$\Phi\equiv0$, hence $\sum_{q=1}^n\gamma_q\Phi_q\equiv 0$. This
contradicts to the fact that the $\Phi_q$ are linearly independent.

It follows that the vectors $\overrightarrow{\Phi}^q$ are linearly
independent and thus they form a Basis of $\mathbb{C}^n$, hence we
can find complex numbers $\gamma_q$, $q=1\ldots n$ such that
$\sum_{q=1}^n \gamma_q\overrightarrow{\Phi}^q=\overrightarrow{f}$.
Defining $f^\parallel:=A\sum_{q=1}^n
\gamma_q\Phi^q\in\mathcal{M}^\parallel$ it follows that
$\left\langle f,A,\Phi_p\right\rangle=\left\langle
f^\parallel,A,\Phi_p\right\rangle$ for any $p=1\ldots n$, i.e.
$f^\bot:=f-f^\parallel\in\mathcal{M}^\bot$.

\item[(b) ``$\subset$''] This proof is equivalent to (a) ``$\subset$''.
Define $\overrightarrow{f}\in\mathbb{R}^n$ by $f_p:=\left\langle
f,A,\Phi_p\right\rangle$ and for any $q=1\ldots n$ a vector
$\overrightarrow{\Phi}^q$ by $\Phi^q_p:=\left\langle \Phi_q,
A,\Phi_p\right\rangle$. Under the assumption that the
$\overrightarrow{\Phi}^q$ are linearly dependent we have the
existence of non trivial $\gamma_q$ such that $\left\langle
\sum_{q=1}^n \gamma_q\Phi_q, A,\sum_{p=1}^n \Phi_p\right\rangle=0$
for any $p=1\ldots n$. Since $\mathcal{N}\cap\mathcal{M}^\bot=\{0\}$
it follows that $\sum_{q=1}^n\gamma_q\Phi_q\equiv0$ which
contradicts to the linear independence of the $\Phi_q$.

It follows that the $\overrightarrow{\Phi}^q$ are linearly
independent, hence we can find complex numbers $\gamma_q$,
$q=1\ldots n$ such that $\sum_{q=1}^n
\gamma_j\overrightarrow{\Phi}^q=\overrightarrow{f}$. Defining
$f^\parallel:=\sum_{q=1}^n \gamma_q\Phi^q\in\mathcal{N}$ it follows
that $\left\langle f,A,\Phi_p\right\rangle=\left\langle
f^\parallel,A,\Phi_p\right\rangle$ for any $p=1\ldots n$, i.e.
$f^\bot:=f-f^\parallel\in\mathcal{M}^\bot$.
\end{description}
$\square$

We now arrive at the main Lemma.

\begin{lemma}\label{haupt}

Let $A\in\mathcal{C}$. Then there exist constants
 $C,C'>0$,  a selfadjoint sesquilinear map $r:\mathcal{N}\times\mathcal{N}\rightarrow\mathbb{C}$
 and a anti-selfadjoint sesquilinear map
 $s:\mathcal{N}\times\mathcal{N}\rightarrow\mathbb{C}$ with $s(\chi,\chi)\neq0$ for all $\chi\in\mathcal{N}$ such that for any
$\mathbf{k}\in\mathbb{R}^3$ with $k<1$, any potential $B$ with $B\in L^1\cap L^\infty$, any normalized
$m^\bot\in\mathcal{M}^\bot$ and any normalized $\Phi,\Psi\in\mathcal{N}$

\begin{itemize}
\item[(a)]
$$|\left\langle \Phi,A,(1-T^{A+B}_{E_k})m^\bot\right\rangle|<C (\|A\|_1+\|B\|_1)(\overline{\lambda}  k+k^2)\;,$$

\item[(b)]

$$\| P^\bot_\mathcal{M} (1-T^{A+B}_{E_k})m^\bot\|_\infty\geq C-C'(\overline{\lambda}  k+k^2+\|B\|_1+\|B\|_\infty)\;,$$

\item[(c)]

$$\| P^\bot_\mathcal{M} (1-T^{A+B}_{E_k})\Phi\|_\infty< C(\overline{\lambda} k+k^2+\|B\|_1+\|B\|_\infty)\;,$$

\item[(d)]

\begin{eqnarray*}
\left\langle \Phi,A, (1-T^{A+B}_{E_k})\Psi\right\rangle&=&
\left\langle \Phi,B,\Psi\right\rangle+r(\Phi,\Psi)
k^2\\&&+ir(\Phi,\lambda(\Psi))
k+s(\Phi,\Psi)k^3+\text{o}(k^3)+\|B\|_1( \overline{\lambda}
\mathcal{O}(k)+\mathcal{O}(k^2))\;.
\end{eqnarray*}

\end{itemize}

\end{lemma}

\noindent\textbf{Proof:}

The proof is given in the Appendix.

$\square$

Using this Lemma we can estimate the inverse of $1-T^{A+B}_{E_k}$.

\begin{lemma}\label{cor}

Let $A\in\mathcal{C}$.  Then there exist constants\\
 $C,C',K,k_0>0$, $C_0,C_1\in\mathbb{R}$, a selfadjoint sesquilinear map $r:\mathcal{N}\times\mathcal{N}\rightarrow\mathbb{C}$
 and an anti-selfadjoint sesquilinear map $s:\mathcal{N}\times\mathcal{N}\rightarrow\mathbb{C}$ with $s(\chi,\chi)\neq0$ for all $\chi\in\mathcal{N}$
 such that for any normalized $\Phi\in\mathcal{N}$, any
$\mathbf{k}\in\mathbb{R}^3$ with $k<k_0$, any potential
$B\in\mathcal{W}_K$ there exists a normalized $\Psi\in\mathcal{N}$
such that
\begin{eqnarray}\label{erstens}
\|P_{\mathcal{N}}^\parallel(1-T^{A+B}_{E_k})^{-1}(A\Phi)\|\leq
C\left(\sup_{\chi\in \mathcal{N},\|\chi\|=1}|\left\langle
\chi,B,\Psi\right\rangle+r(\chi,\Psi) k^2- ik \langle
\chi,A,\lambda(\Psi) \rangle +s(\chi,\Psi)k^3|
 \right)^{-1}\nonumber
\end{eqnarray}
and
\begin{eqnarray}\label{zweitens}
&&\| P_{\mathcal{N}}^\bot(1-T^{A+B}_{E_k})^{-1}(A\Phi)\|_\infty\leq
C (\overline{\lambda}
k+k^2+\|B\|_1+\|B\|_\infty)\nonumber\\&&\left(\sup_{\chi\in
\mathcal{N},\|\chi\|=1}|\left\langle
\chi,B,\Psi\right\rangle+r(\chi,\Psi) k^2-ik \langle
\chi,A,\lambda(\Psi) \rangle+s(\chi,\Psi)k^3|
 \right)^{-1}\;.
\end{eqnarray}
Furthermore we have that for any normalized
$m^\bot\in\mathcal{M}^\bot$ there exists a normalized
$\Psi\in\mathcal{N}$  such that
\begin{eqnarray}\label{drittens}
&&\|
P_{\mathcal{N}}^\parallel(1-T^{A+B}_{E_k})^{-1}m^\bot\|_\infty\\&&\nonumber\leq
C|\overline{\lambda} k+k^2|\left(\sup_{\chi\in
\mathcal{N},\|\chi\|=1}|\left\langle
\chi,B,\Psi\right\rangle+r(\chi,\Psi) k^2-ik \langle
\chi,A,\lambda(\Psi) \rangle+s(\chi,\Psi)k^3|
 \right)^{-1}
\end{eqnarray}
and
\begin{equation}\label{viertens}
\| P_{\mathcal{N}}^\bot(1-T^{A+B}_{E_k})^{-1}m^\bot\|_\infty\leq
C\;.\end{equation}

\end{lemma}

\noindent\textbf{Proof:} Let $A\in\mathcal{C}$. Choose $k_0$ and $K$
(there will be further restrictions on $k_0$ and $K$ below, so the
final $k_0$ and $K$ may at the end be smaller) such that there
exists a $C>0$ such that
\begin{equation}\label{glei1}
\| P^\bot_\mathcal{M} (1-T^{A+B}_{E_k})h^\bot\|_\infty\geq C
\end{equation}
for any $h^\bot\in\mathcal{M}^\bot$, any
$\mathbf{k}\in\mathbb{R}^3$ with $k<k_0$ and any potential
$B\in\mathcal{W}_K$ (in view of Lemma \ref{haupt} (b) such a
choice is possible).

Then (using Lemma \ref{haupt} (a), (c) and (d)) one can find a
constant $C>0$ such that for any $\mathbf{k}\in\mathbb{R}^3$ with
$k<k_0$, any potential $B\in\mathcal{W}_K$ (i.e. bounded $\|B\|_1$)
and any normalized
$\widetilde{\Phi}\in\mathcal{N},m^\bot\in\mathcal{M}^\bot$
\begin{equation}\label{glei2}
\sup_{\chi\in \mathcal{N},\|\chi\|=1}|\left\langle
\chi,A,(1-T^{A+B}_{E_k})m^\bot\right\rangle|<C (\overline{\lambda}
k+k^2)=:t_1\;,
\end{equation}
\begin{equation}\label{glei3}\| P^\bot_\mathcal{M} (1-T^{A+B}_{E_k})\Phi\|_\infty< C
(\overline{\lambda} k+k^2+\|B\|_1+\|B\|_\infty)=:t_2
\end{equation}
and
\begin{eqnarray*}
&&\hspace{-1cm}\sup_{\chi\in \mathcal{N},\|\chi\|=1}\left\langle
\chi,A, (1-T^{A+B}_{E_k})\Phi\right\rangle\\&=&
 \sup_{\chi\in \mathcal{N},\|\chi\|=1}|\left\langle \chi,B,\Phi\right\rangle-ik \langle
\chi,A,\lambda(\Phi)
\rangle+k^2r(\chi,\Phi)+k^3s(\chi,\Phi)|\\&&+\text{o}(k^3)+\mathcal{O}(\|B\|_1)(\overline{\lambda}
\mathcal{O}(k)+\mathcal{O}(k^2))\;.\nonumber
\end{eqnarray*}
Next we will show that the first summand will suffice for our
estimates, i.e. that there exists a constant $C>0$ such that
\begin{eqnarray}\label{glei4}
&&\hspace{-1cm}\sup_{\chi\in \mathcal{N},\|\chi\|=1}|\left\langle
\chi,A,
(1-T^{A+B}_{E_k})\Phi\right\rangle|\\\nonumber&\geq&C\sup_{\chi\in
\mathcal{N},\|\chi\|=1}\left|\left\langle
\chi,B,\Phi\right\rangle-ik \langle \chi,A,\lambda(\Phi)
\rangle+k^2r(\chi,\Phi)+k^3s(\chi,\Phi)\right|=:t_3\;.
\end{eqnarray}
Therefore we have to show that for sufficiently small $K,k_0$:
\begin{eqnarray}\label{glei4b}t_3\gg\text{o}(k^3)+\mathcal{O}(\|B\|_1)(\overline{\lambda}
\mathcal{O}(k)+\mathcal{O}(k^2))
\end{eqnarray}
which we will do next.

We will prove (\ref{glei4b}) for $\overline{\lambda}=1$,
$\overline{\lambda} =0$ and $\|B\|_1=\mathcal{O}(k)$ and
$\overline{\lambda} =0$ and $\|B\|_1\gg k$ separately.
\begin{itemize}
    \item[$1^{st}$ Case:] Assume that $\overline{\lambda}=1$. Then the leading order of $t_3$ is obviously greater than or equal to
    $\left\langle
\Psi,B,\Psi\right\rangle-ik \langle \Psi,A,\lambda(\Psi)\rangle$.
The first summand is real, the second summand is (see
(\ref{wirdnull}))
$$-ik\int\int\Psi^t(\mathbf{x})A(\mathbf{x}) (1+\beta)A(\mathbf{y})\Psi(\mathbf{y})d^3y d^3x=-ik\lambda^t(\Psi)(1+\beta)\lambda(\Psi)$$
and - since $\beta$ is selfadjoint - imaginary (and not equal to
zero). Hence there exists a $C>0$ such that $\left\langle
\Psi,B,\Psi\right\rangle-ik \langle \Psi,A,\lambda(\Psi)\rangle>Ck$
and (\ref{glei4b}) holds.
    \item[$2^{nd}$ Case:] Assume that $\overline{\lambda} =0$ and $\|B\|_1=\mathcal{O}(K^{-\frac{1}{2}}k)$. Similar as above there exists a $C>0$ such that
$\left\langle
\Psi,B,\Psi\right\rangle-k^2r(\chi,\Phi)-k^3s(\chi,\Phi)>Ck^3$.
Since in this case $\|B\|_1\mathcal{O}(k^2)=\mathcal{O}(k^3)$
equation (\ref{glei4b}) holds.
    \item[$3^{rd}$ Case:] Assume that $\overline{\lambda} =0$ and $\|B\|_1\gg K^{-\frac{1}{2}}k$.
    Since $B\in\mathcal{W}_K$ it follows that $\left\langle \Psi,B,\Psi\right\rangle\gg k^2$, hence
$\left\langle \Psi,B,\Psi\right\rangle-k^2r(\chi,\Phi)\gg k^2$ and
(\ref{glei4b}) holds.

\end{itemize}

We next prove (\ref{erstens}) and (\ref{zweitens}).
 We define
\begin{eqnarray}\label{hpar}\omega \Psi&:=&P_{\mathcal{N}}^\parallel
(1-T^{A+B}_{E_k})^{-1}(A\Phi)
\\
\label{hsenk} h^\bot(\mathbf{k},B)&:=&P_{\mathcal{N}}^\bot
(1-T^{A+B}_{E_k})^{-1}(A\Phi)\;,
\end{eqnarray}
with $\omega>0$ and  $\|\Psi\|_\infty=1$.

It follows that
$$(1-T^{A+B}_{E_k})(\omega\Psi+h^\bot(\mathbf{k},B))=A\Phi\;,$$
hence
\begin{eqnarray*}
\sup_{\chi\in
\mathcal{N},\|\chi\|=1}|\left\langle\chi,A,(1-T^{A+B}_{E_k})(\omega\Psi+h^\bot(\mathbf{k},B))\right\rangle|=\sup_{\chi\in
\mathcal{N},\|\chi\|=1}|\left\langle\chi,A,A\Phi\right\rangle|:=C_\Phi
\end{eqnarray*}
and
$$P_{\mathcal{M}}^\bot(1-T^{A+B}_{E_k})(\omega\Psi+h^\bot(\mathbf{k},B))=0\;.$$
Using (\ref{glei4}) and (\ref{glei2}) we get
\begin{eqnarray*}
t_3 \omega
<t_1 \|h^\bot(\mathbf{k},B)\|_\infty+C_\Phi\;,
\end{eqnarray*}
using (\ref{glei3}) and (\ref{glei1}) we get
\begin{eqnarray}\label{fuer2}
 t_2|
\omega|\geq C\| h^\bot(\mathbf{k},B)\|_\infty\;,
\end{eqnarray}
hence
\begin{eqnarray*}
t_3\omega
&<&t_1  \frac{t_2}{C}\omega+C_\Phi
\\
t_3 \omega
-t_1  \frac{t_2}{C}\omega&<&C_\Phi
\\
\omega&<&C_\Phi(t_3-\frac{t_1t_2}{C})^{-1} \;.
\end{eqnarray*}
Note, that  $C_\Phi$ is bounded uniformly in normalized $\Phi$. To
get (\ref{erstens}) it is left to show that for small enough
$k_0,K$
\begin{equation}\label{gleichung1}\frac{t_1 t_2}{C}<\frac{t_3}{2}\end{equation}
uniform in $k<k_0$ and $B\in\mathcal{W}_K$.

We will prove (\ref{gleichung1}) for $\overline{\lambda}=1$,
$\overline{\lambda} =0$ and
$\|B\|_1+\|B\|_\infty=\mathcal{O}(\sqrt{K}k)$ and
$\overline{\lambda} =0$ and $\|B\|_1+\|B\|_\infty\gg \sqrt{K}k$
separately.
\begin{itemize}
    \item[$1^{st}$ Case:] Assume that $\overline{\lambda}=1$.
Then we have that $t_1 t_2$ is of order $k(\overline{\lambda}
k+k^2+\|B\|_1+\|B\|_\infty)$ and $|t_3|$ is of order $k$ (see
above). Hence for $K$ small enough (i.e. $\|B\|_1$ and
$\|B\|_\infty$ small enough) and $k_0$ small enough
(\ref{gleichung1}) follows.

\item[$2^{nd}$ Case:] Assume that $\overline{\lambda} =0$ and $\|B\|_1+\|B\|_\infty=\mathcal{O}(\sqrt{K}k)$. Then we have
that $t_3$ is of order $k^3$ and $t_1 t_2$ is of order $k^2(k^2+\|B\|_1+\|B\|_\infty)$. Hence for small enough
$K$ (\ref{gleichung1}) follows.

\item[$3^{rd}$ Case:] Assume that $\overline{\lambda} =0$ and $\|B\|_1+\|B\|_\infty\gg \sqrt{K}k$, i.e. $$\sup_{\chi\in
\mathcal{N},\|\chi\|=1}|\left\langle \chi,B,\Phi\right\rangle|\gg
k^2\;.$$ It follows that $\sup_{\chi\in
\mathcal{N},\|\chi\|=1}|\left\langle
\chi,B,\Phi\right\rangle-k^2r(\chi,\Phi)\gg k^2$, hence $t_3\gg k^2$
and (\ref{gleichung1}) follows.

\end{itemize}

In view of (\ref{erstens}) and (\ref{fuer2}) we have that
$$\| h^\bot(\mathbf{k},B)\|_\infty\leq \frac{t_2}{C t_3}\;,$$
which is - in view of (\ref{glei3}), (\ref{glei4}) and
(\ref{hsenk}) - exactly (\ref{zweitens}).

(\ref{drittens}) and (\ref{viertens}) can be verified in a similar way as (\ref{erstens}) and (\ref{zweitens}).
We define
\begin{eqnarray}\label{hpar2}\omega \Psi&:=&P_{\mathcal{N}}^\parallel
(1-T^{A+B}_{E_k})^{-1}(m^\bot)
\\\label{hsenk2}
h^\bot(\mathbf{k},B)&:=&P_{\mathcal{N}}^\bot
(1-T^{A+B}_{E_k})^{-1}(m^\bot)\;,
\end{eqnarray}
with $\omega>0$ and  $\|\Psi\|_\infty=1$.

It follows that
$$(1-T^{A+B}_{E_k})(\omega \Psi+h^\bot(\mathbf{k},B))=m^\bot\;,$$
hence
$$\sup_{\chi\in \mathcal{N},\|\chi\|=1}|\left\langle\chi,A,(1-T^{A+B}_{E_k})(\omega \Psi+h^\bot(\mathbf{k},B))\right\rangle|=0$$
and
$$\|P_{\mathcal{M}}^\bot(1-T^{A+B}_{E_k})(\omega \Psi+h^\bot(\mathbf{k},B))\|_\infty=1\;.$$

Using (\ref{glei4}) and (\ref{glei2}) we get
\begin{eqnarray}\label{40}
t_3 |\omega|
<t_1 \|h^\bot(\mathbf{k},B)\|_\infty\;,
\end{eqnarray}
using (\ref{glei3}) and (\ref{glei1}) we get
\begin{eqnarray}\label{fuer2b}
1-t_2|\omega|\geq C\| h^\bot(\mathbf{k},B)\|_\infty
\end{eqnarray}
hence
\begin{eqnarray}\label{hierbb}
C\| h^\bot(\mathbf{k},B)\|_\infty\leq1-\frac{t_1t_2}{t_3} \|h^\bot(\mathbf{k},B)\|_\infty\;.
\end{eqnarray}

In view of (\ref{gleichung1}) $\frac{t_1t_2}{t_3}<\frac{1}{2}C$. It follows that $\|
h^\bot(\mathbf{k},B)\|_\infty$ is of order one, which is exactly (\ref{viertens}).

In view of (\ref{viertens}) and (\ref{40}) we have that
$$|\omega|\leq \frac{t_1}{C t_3}$$
which is - in view of (\ref{glei3}), (\ref{glei4}) and
(\ref{hsenk2}) - exactly (\ref{zweitens}).

$\square$

The Lemma can be written in a much nicer way, separating the
different cases $\overline{\lambda}=0$ and $\overline{\lambda}=1$.
For $\overline{\lambda}=0$

\begin{corollary}\label{cor2a} 

Let  $A\in\mathcal{C}$ with $\overline{\lambda}=0$ (i.e.
$\mathcal{N}\subset L^2$). Then there exist constants
 $C,K,k_0>0$ and a selfadjoint linear map $\widehat{R}:\mathcal{N}\to\mathcal{N}$
 such that for any normalized $\Phi\in\mathcal{N}$, any
$\mathbf{k}\in\mathbb{R}^3$ with $k<k_0$, any potential
$B\in\mathcal{W}_K$
\begin{eqnarray}\label{erstensa}
\|P_{\mathcal{N}}^\parallel(1-T^{A+B}_{E_k})^{-1}(A\Phi)\|\leq
C\left(\inf_{\Psi\in \mathcal{N},\|\Psi\|=1}\left\|
\left(P_{\mathcal{N}}^\parallel B+
\widehat{R}k^2\right)\Psi\right\|+k^3
 \right)^{-1}
\end{eqnarray}
and
\begin{eqnarray}\label{zweitensa}
\| P_{\mathcal{N}}^\bot(1-T^{A+B}_{E_k})^{-1}(A\Phi)\|_\infty\leq C
(k^2+\|B\|_1+\|B\|_\infty)\left(\inf_{\Psi\in
\mathcal{N},\|\Psi\|=1}\left\| \left(P_{\mathcal{N}}^\parallel B+
\widehat{R}k^2\right)\Psi\right\|+k^3
 \right)^{-1}\;.
\end{eqnarray}
Furthermore we have for any normalized $m^\bot\in\mathcal{M}^\bot$
\begin{eqnarray}\label{drittensa}
\|
P_{\mathcal{N}}^\parallel(1-T^{A+B}_{E_k})^{-1}m^\bot\|_\infty\leq
Ck^2\left(\inf_{\Psi\in \mathcal{N},\|\Psi\|=1}\left\|
\left(P_{\mathcal{N}}^\parallel B+
\widehat{R}k^2\right)\Psi\right\|+k^3
 \right)^{-1}
\end{eqnarray}
and
\begin{equation}\label{viertensa}
\| P_{\mathcal{N}}^\bot(1-T^{A+B}_{E_k})^{-1}m^\bot\|_\infty\leq
C\;.\end{equation}
\end{corollary}

\noindent\textbf{Proof:} The Corollary follows directly from Lemma
\ref{cor}. Note, that we consider the case $\mathcal{N}\subset L^2$,
i.e. there exists a selfadjoint linear map
$\widehat{R}:\mathcal{N}\to\mathcal{N}$ and a antiselfadjoint linear
map $\widehat{S}:\mathcal{N}\to\mathcal{N}$ such that
$\langle\Phi,\widehat{R}\chi\rangle=r(\Phi,\chi)$ and
$\langle\Phi,\widehat{S}\chi\rangle=s(\Phi,\chi)$ for $r$ and $s$
coming from the Lemma. Recall that $s(\Phi,\Phi)\neq0$ for all
$\Phi\in\mathcal{N}\backslash\{0\}$, hence $
\left\langle\Phi,S\Phi\right\rangle\neq0$ for all
$\Phi\in\mathcal{N}\backslash\{0\}$

Using this and $\overline{\lambda}=0$ we have
\begin{eqnarray}\label{corgleich}\sup_{\chi\in \mathcal{N},\|\chi\|=1}|\left\langle
\chi,B,\Psi\right\rangle+r(\chi,\Psi) k^2+s(\chi,\Psi)k^3| =
\sup_{\chi\in \mathcal{N},\|\chi\|=1}|\langle
\chi,B+\widehat{R}k^2+\widehat{S}k^3,\Psi\rangle|\;.
\end{eqnarray}
Note, that
\begin{eqnarray*}
\sup_{\chi\in \mathcal{N},\|\chi\|=1}|\langle
\chi,B+\widehat{R}k^2+\widehat{S}k^3,\Psi\rangle|&\geq&|\langle
\Psi,B+\widehat{R}k^2+\widehat{S}k^3,\Psi\rangle|\;.
\end{eqnarray*}
Since $B$ and $\widehat{R}$ are selfadjoint and $\widehat{S}$ is
anti-selfadjoint, the first two summands are real, the last is
imaginary. Furthermore we have that $
 \langle\Phi,S\Phi \rangle\neq0$ for all
$\Phi\in\mathcal{N}\backslash\{0\}$. Hence there exists a constants
$C \in\mathbb{R}\backslash\{0\}$ such that
\begin{equation}\label{sop}\sup_{\chi\in \mathcal{N},\|\chi\|=1}| \langle
\chi,B+\widehat{R}k^2+\widehat{S}k^3,\Psi \rangle|\geq C k^3\;.
\end{equation}
Furthermore we have that
\begin{eqnarray*}
&&\sup_{\chi\in \mathcal{N},\|\chi\|=1}| \langle
\chi,B+\widehat{R}k^2+\widehat{S}k^3,\Psi \rangle|
\geq\sup_{\chi\in \mathcal{N},\|\chi\|=1}| \langle
\chi,B+\widehat{R}k^2,\Psi \rangle| -k^3\|\widehat{S}\|^{op}\;,
\end{eqnarray*}
hence with (\ref{sop}) there exists a $C>0$ such that
\begin{eqnarray*}
C\sup_{\chi\in \mathcal{N},\|\chi\|=1}| \langle
\chi,B+\widehat{R}k^2+\widehat{S}k^3,\Psi \rangle|
&\geq&\sup_{\chi\in \mathcal{N},\|\chi\|=1}| \langle
\chi,B+\widehat{R}k^2,\Psi \rangle| +k^3
\\&=&\left\|\left(P_{\mathcal{N}}^\parallel B+
\widehat{R}k^2\right)\Psi\right\|+k^3
\\&\geq&
\inf_{\widetilde{\Psi}\in \mathcal{N},\|\widetilde{\Psi}\|=1}\left\|
\left(P_{\mathcal{N}}^\parallel B+
\widehat{R}k^2\right)\widetilde{\Psi}\right\|+k^3\;.
\end{eqnarray*}
Using this formula and $\overline{\lambda}=0$ in Lemma \ref{cor} the
Corollary follows.

$\square$  For $\overline{\lambda}=1$ we have for Lemma \ref{cor}
\begin{corollary}\label{cor2b} 
Let $A\in\mathcal{C}$ and $\overline{\lambda}=1$ (i.e.
$\mathcal{N}\cap L^2=\emptyset$). Then there exist constants
 $C,K,k_0>0$ such that for any normalized $\Phi\in\mathcal{N}$, any
$\mathbf{k}\in\mathbb{R}^3$ with $k<k_0$ and any potential
$B\in\mathcal{W}_K$
\begin{eqnarray}\label{erstensb}
\|P_{\mathcal{N}}^\parallel(1-T^{A+B}_{E_k})^{-1}(A\Phi)\|\leq
C\left(\inf_{\Psi\in \mathcal{N},\|\Psi\|=1}\left|\left\langle
\Psi,B,\Psi\right\rangle\right|+k
 \right)^{-1}
\end{eqnarray}
and
\begin{eqnarray}\label{zweitensb}
\| P_{\mathcal{N}}^\bot(1-T^{A+B}_{E_k})^{-1}(A\Phi)\|_\infty\leq C
( k+\|B\|_1+\|B\|_\infty)\left(\inf_{\Psi\in
\mathcal{N},\|\Psi\|=1}\left|\left\langle
\Phi,B,\Phi\right\rangle\right|+k
 \right)^{-1}\;.
\end{eqnarray}
Furthermore we have for any normalized $m^\bot\in\mathcal{M}^\bot$
\begin{eqnarray}\label{drittensb}
\|
P_{\mathcal{N}}^\parallel(1-T^{A+B}_{E_k})^{-1}m^\bot\|_\infty\leq
Ck\left(\inf_{\Psi\in \mathcal{N},\|\Psi\|=1}\left|\left\langle
\Phi,B,\Phi\right\rangle\right|+k
 \right)^{-1}
\end{eqnarray}
and
\begin{equation}\label{viertensb}
\| P_{\mathcal{N}}^\bot(1-T^{A+B}_{E_k})^{-1}m^\bot\|_\infty\leq
C\;.\end{equation}

\end{corollary}

\noindent\textbf{Proof:} The proof is as the proof of Corollary
\ref{cor2a} above, using that $B$ is selfadjoint and $q$ is
anti-selfadjoint.

$\square$

Next we show, how these corollaries imply the Theorem. First recall
(\ref{LSE3})
$$(1-
T^{A+B}_{E_k})\phi(A+B,j,\mathbf{k},\cdot)=\chi(j,\mathbf{k},\cdot)\;.$$
Defining
\begin{equation}\label{defg}g(A+B,j,\mathbf{k},\cdot)=T^{A+B}_{E_k}\chi(j,\mathbf{k},\cdot)\end{equation}
and
\begin{equation}\label{defzeta}\zeta(A+B,j,\mathbf{k},\cdot)=\phi(A+B,j,\mathbf{k},\cdot)-\chi(j,\mathbf{k},\cdot)\end{equation}
it follows that
\begin{eqnarray}\label{geq}\zeta(A+B,j,\mathbf{k},\cdot)&=&-(1-
T^{A+B}_{E_k})^{-1}g(A+B,j,\mathbf{k},\cdot)\nonumber\\&=&-(1-
T^{A+B}_{E_k})^{-1}P^\parallel_\mathcal{M}g(A+B,j,\mathbf{k},\cdot)\nonumber\\&&-(1-
T^{A+B}_{E_k})^{-1}P^\bot_\mathcal{M}g(A+B,j,\mathbf{k},\cdot)\;.
\end{eqnarray}
\noindent\textbf{Proof of Theorem \ref{theo}:} Below we shall show
that that for $\overline{\lambda} =0$
\begin{equation}\label{leftto}\|P^\parallel_\mathcal{M}g(A+B,j,\mathbf{k},\cdot)\|_\infty< C \left(k+\|B\|_1\right)\;.
\end{equation}
Defining \begin{equation}\label{defphib}
\widetilde{\Phi}^B_{j,\mathbf{k}}:=P^\parallel_\mathcal{N}\zeta(A+B,j,\mathbf{k},\cdot)
\end{equation} and using Corollary \ref{cor2a} in (\ref{geq}) one gets
\begin{eqnarray}\label{redef1}\|\widetilde{\Phi}^B_{j,\mathbf{k}}\|&\leq&
\|P^\parallel_\mathcal{N}(1-
T^{A+B}_{E_k})^{-1}P^\parallel_\mathcal{M}g(A+B,j,\mathbf{k},\cdot)\|\nonumber+\|P^\parallel_\mathcal{N}(1-
T^{A+B}_{E_k})^{-1}P^\bot_\mathcal{M}g(A+B,j,\mathbf{k},\cdot)\|
\nonumber\\&\leq&C\left(k+\|B\|_1\right)\left(\inf_{\Psi\in
\mathcal{N},\|\Psi\|=1}\left\| \left(P_{\mathcal{N}}^\parallel B+
\widehat{R}k^2\right)\Psi\right\|+k^3
 \right)^{-1}\nonumber\\&&+Ck^2\left(\inf_{\Psi\in \mathcal{N},\|\Psi\|=1}\left\|
\left(P_{\mathcal{N}}^\parallel B+
\widehat{R}k^2\right)\Psi\right\|+k^3
 \right)^{-1}
 \nonumber\\&\leq&C\left(k+\|B\|_1\right)\left(\inf_{\Psi\in \mathcal{N},\|\Psi\|=1}\left\|
\left(P_{\mathcal{N}}^\parallel B+
\widehat{R}k^2\right)\Psi\right\|+k^3
 \right)^{-1}
\end{eqnarray}
and in view of (\ref{defzeta})
\begin{eqnarray}\label{redef2}&&\|\phi(A+B,j,\mathbf{k},\cdot)-\widetilde{\Phi}^B_{j,\mathbf{k}}\|
\nonumber\\\nonumber&\leq&\|\chi(j,\mathbf{k},\cdot)\|_\infty+\|P^\bot_\mathcal{N}\zeta(A+B,j,\mathbf{k},\cdot)\|_\infty
\nonumber\\&\leq&1+
\|P^\bot_\mathcal{N}(1-
T^{A+B}_{E_k})^{-1}g(A+B,j,\mathbf{k},\cdot)\|+\|P^\bot_\mathcal{N}(1-
T^{A+B}_{E_k})^{-1}P^\bot_\mathcal{M}g(A+B,j,\mathbf{k},\cdot)\|
\nonumber\\&\leq&1+Ck (k^2+\|B\|_1+\|B\|_\infty)\left(\inf_{\Psi\in
\mathcal{N},\|\Psi\|=1}\left\| \left(P_{\mathcal{N}}^\parallel B+
\widehat{R}k^2\right)\Psi\right\|+k^3
 \right)^{-1}+
C
\nonumber\\&\leq&C+Ck\left(\|B\|_1+\|B\|_\infty\right)\left(\inf_{\Psi\in
\mathcal{N},\|\Psi\|=1}\left\| \left(P_{\mathcal{N}}^\parallel B+
\widehat{R}k^2\right)\Psi\right\|+k^3
 \right)^{-1}\;.
\end{eqnarray}

These formulas imply the Theorem, it is left to verify
(\ref{leftto}). Using the equivalence of all norms in the finite
dimensional space $\mathcal{M}$ we have that there exists a $C>0$
and a normalized $\Phi\in\mathcal{N}$ such that
\begin{eqnarray*}
\|P^\parallel_\mathcal{M}g(A+B,j,\mathbf{k},\cdot)\|_\infty&\leq&C\left\langle
\Phi,A,g(A+B,j,\mathbf{k},\cdot)\right\rangle\;.
\end{eqnarray*}
In view of (\ref{defg}) we have
\begin{eqnarray*}
&&|\left\langle
\Phi,A,g(A+B,j,\mathbf{k},\cdot)\right\rangle|\\&&=|\left\langle
\Phi,A,T^{A+B}_{1}\chi(j,\mathbf{k},\cdot)\right\rangle+\left\langle
\Phi,A,(T^{A+B}_{E_k}-T^{A+B}_{1})\chi(j,\mathbf{k},\cdot)\right\rangle|
\\&&\leq|\left\langle \Phi,A,T^{A+B}_{1}\chi(j,\mathbf{k},\cdot)\right\rangle|+|\left\langle
(T^{A}_{E_k}-T^{A}_{1})\Phi,A+B,\chi(j,\mathbf{k},\cdot)\right\rangle|
\\&&\leq|\left\langle \Phi,A,g(A+B,j,0,\cdot)\right\rangle|+|\left\langle \Phi,A,T^{A+B}_{1}(\chi(j,\mathbf{k},\cdot)-\chi_1)\right\rangle|\\&&+\|
(T^{A}_{E_k}-T^{A}_{1})\Phi\|_\infty\space\|(A+B)\chi(j,\mathbf{k},\cdot)\|_1\;.
\end{eqnarray*}
Remember, that
$\chi(j,\mathbf{k},\mathbf{c})=e^{i\mathbf{k}\cdot\mathbf{x}}$
multiplied with some ($k$-dependent) four-spinor. Hence
$\chi(j,\mathbf{k},x\mathbf{\cdot})-\chi(j,0,\mathbf{x})$ is of
order $k(1+x)$, thus the second summand is of order $k$. In view of
Lemma \ref{hauptalles} (d) using that $\chi(j,\mathbf{k},\cdot)$ is
normalized, the third summand is of order $(\overline{\lambda}
k+k^2)$. It suffices to prove that if $\overline{\lambda} =0$
$$\left\langle
\Phi,A,g(A+B,j,0,\cdot)\right\rangle=\mathcal{O}(\|B\|_1)\;.$$
Therefore we use that $\chi(j,0,\cdot)$ is a generalized
eigenfunction of the free Dirac equation with energy $1$, i.e.
$(1-\beta)\chi(j,0,\cdot)=0$ and thus
$(1+\beta)\chi(j,0,\cdot)=2\chi(j,0,\cdot)$. This (\ref{symm}),
(\ref{defg}) and (\ref{wirdnull}) yields
\begin{eqnarray*}
\left\langle\Phi,A, g(A+B,j,0,\cdot)\right\rangle&=&\left\langle
\Phi,A,T^{A+B}_1\chi(j,0,\cdot)\right\rangle
\\&=&\frac{1}{2}\left\langle T^A_1\Phi,A+B,(1+\beta)\chi(j,0,\cdot)\right\rangle
\\&=&\frac{1}{2}\left\langle \int \Phi A(1+\beta)d^3x,A+B, \chi(j,0,\cdot)\right\rangle
\\&&+\frac{1}{2}\left\langle\int \Phi
B(1+\beta)d^3x,A+B,\chi(j,0,\cdot)\right\rangle
\\&=&\frac{1}{2}\left\langle\lambda(\Phi),A+B,\chi(j,0,\cdot)\right\rangle
+\frac{1}{2}\left\langle\int \Phi B(1+\beta)d^3x,A+B,
\chi(j,0,\cdot)\right\rangle
\\&\leq&C\|B\|_1\;.
\end{eqnarray*}

$\square$

Theorem \ref{theob} follows with (\ref{defzeta}) and using Corollary
\ref{cor2b} in (\ref{geq}).

\section{$k$-Derivatives}\label{sectionkder}

Next we will estimate the $k$-derivatives of the solutions of
(\ref{LSE3}) assuming that $A$ and $B$ are compactly supported. The
results of this section play an important role for the estimate of
wave function decay (see \cite{diss} and \cite{pdneu}) via
stationary phase method.

For ease of writing we define
\begin{eqnarray*}\alpha:=1+\left(k+\|B\|_1\right)\left(\inf_{\Phi\in
\mathcal{N},\|\Phi\|=1}\left\| \left(P_{\mathcal{N}}^\parallel B+
\widehat{R}k^2\right)\Phi\right\|+k^3
 \right)^{-1}\;.\end{eqnarray*}

Heuristically deriving (\ref{LSE3}) with respect to $k$ yields
$\partial_{k}\phi(A+B,j,\mathbf{k},\mathbf{x})$. We denote the
function we get by this formal method by
$\dot\phi(A+B,j,\mathbf{k},\mathbf{x})$.
\begin{eqnarray}\label{defder}
(1-
T^{A+B}_{E_k})\dot\phi(A+B,j,\mathbf{k},\cdot)=\partial_k\chi_\mathbf{k}+(\partial_k
T^{A+B}_{E_k})\phi(A+B,j,\mathbf{k},\cdot)=:f_1\;.
\end{eqnarray}
Similarly as above one defines
$$g^1:=\mu T^{A+B}_{E_k} f^1\;\;\text{and }\zeta^1:=\dot{\phi}-f^1$$
to get
\begin{equation}\label{ableitung}
\zeta^1(A+B,j,\mathbf{k},\cdot)=-(1-
T^{A+B}_{E_k})^{-1}g^1(A+B,j,\mathbf{k},\cdot)\;.
\end{equation}
In \cite{pickl} it is shown that (\ref{ableitung}) has a unique
solution and that in fact $\dot{\phi}=\partial_{k}\phi$.

Now $\dot{\phi}$ is controllable via $\zeta^1$ using
(\ref{ableitung}) in a similar way as we controlled
$\phi(A+B,j,\mathbf{k},\cdot)$ above (c.f. (\ref{geq})). Let us
heuristically estimate $\|\zeta^1(A+B,j,\mathbf{k},\cdot)\|_\infty$
for $\overline{\lambda}=0$ to make the result clear, a rigorous
treatment (which is in fact ``not far'' from this heuristics) shall
be given below in more generality (i.e. for higher derivatives,
also). Recall that $\|\phi(A+B,j,\mathbf{k},\cdot)\|_\infty\leq
C\alpha$ for appropriate $C<\infty$. Since
\begin{eqnarray*}
(\partial_k T^{A+B}_{E_k})\phi(A+B,j,\mathbf{k},\cdot)&=&(\partial_k
T^{A}_{E_k})\phi(A+B,j,\mathbf{k},\cdot)
+(\partial_k T^{B}_{E_k})\phi(A+B,j,\mathbf{k},\cdot)
\\&=&[\partial_k
T^{A}_{E_k}\phi(A+B,j,\mathbf{k},\cdot)]_{k=0}+\left(\mathcal{O}(k)+\mathcal{O}(\|B\|_1)\right)\|\phi(A+B,j,\mathbf{k},\cdot)\|_\infty\;.
\end{eqnarray*}
In view of (\ref{dktphi}) the first summand is zero for
$\overline{\lambda}=0$ and $g^1$ is bounded from above by
$C(k+\|B\|_1)\alpha$ . Using as above Corollary \ref{cor2a} we get
that $\partial_{k}\phi\leq C\alpha^2$ for appropriate $C$.

Heuristically one can treat the higher derivatives similarly, hence
we have
\begin{thm}\label{lambda0der}
Let  $A\in\mathcal{C}$ with $\overline{\lambda}=0$ (i.e.
$\mathcal{N}\subset L^2$). Then there exist constants
 $C,K,k_0>0$ and a selfadjoint linear map $\widehat{R}:\mathcal{N}\to\mathcal{N}$
 such that for any $m\in\mathbb{N}_0$ there exist $C_m<\infty$ such
 that for any
$\mathbf{k}\in\mathbb{R}^3$ with $k<k_0$, $j=1,2$, any potential
$B\in\mathcal{W}_K$ there exists a
$\Phi^B_{j,\mathbf{k}}\in\mathcal{N}$ with
\begin{equation*}\|(1+x)^{-m}\partial_k^m\phi(A+B,j,\mathbf{k},\cdot)\|_\infty\leq
C_m\left(k^{-m}+\alpha^{m+1}\right)\;.
\end{equation*}
\end{thm}
\noindent\textbf{Proof:} We repeat the procedure above which gave us
the defining equation for $\partial_k\phi$ (i.e. (\ref{defder})) for
the higher derivatives. We get formally
\begin{equation}\label{genbg}
\partial_k^m\left((1-T^{A+B}_{E_k})\phi(A+B,j,\mathbf{k},\cdot)\right)=\partial_k^m \chi(j,\mathbf{k},\cdot)\;,
\end{equation}
hence
\begin{eqnarray}\label{genbg2}
\left((1-
T^{A+B}_{E_k})\phi^{(m)}(A+B,j,\mathbf{k},\cdot)\right)=\partial_k^m
\chi(j,\mathbf{k},\cdot)
-\sum_{l=1}^m\left(%
\begin{array}{c}
  m \\
  l \\
\end{array}%
\right)
\partial_k^lT^{A+B}_{E_k}\partial_k^{(m-l)}\phi(A+B,j,\mathbf{k},\cdot)\;.
\end{eqnarray}
Defining
\begin{equation}\label{deff}
f^{m}(A+B,j,\mathbf{k},\cdot):=\partial_k^m \chi(j,\mathbf{k},\cdot)- \sum_{l=1}^m\left(%
\begin{array}{c}
  m \\
  l \\
\end{array}%
\right)
\partial_k^lT^{A+B}_{E_k}\partial_k^{(m-l)}\phi(A+B,j,\mathbf{k},\cdot)\;,
\end{equation}
\begin{equation}\label{ggen}
g^m(A+B,j,\mathbf{k},\cdot):=
T^{A+B}_{E_k}f^m(A+B,j,\mathbf{k},\cdot)\end{equation} and
$$\zeta^{(m)}(A+B,j,\mathbf{k},\cdot):=\phi^{(m)}(A+B,j,\mathbf{k},\cdot)-f^m(A+B,j,\mathbf{k},\cdot)\;,$$
it follows that
\begin{equation}\label{gengeq}(1-T^{A+B}_{E_k})\zeta^{(m)}(A+B,j,\mathbf{k},\cdot)=-g^m(A+B,j,\mathbf{k},\cdot)\;.
\end{equation}
Again \cite{pickl} shows that the formal differentiations yield the
right functions, i.e. $\phi^{m}=\partial_k^m\phi$.

First we will show inductively that there exist $C_m<\infty$ and
$\Phi_m\in\mathcal{N}$ such that
\begin{eqnarray}\label{ind1}
\|(1+x)^{-m+1} f^m(A+B,j,\mathbf{k},\cdot)\|_{\infty}&\leq& C_m\alpha(k^{-1}+\alpha)^{m-1}
%
%
%
\\\label{ind4}\|\Phi_m\|&\leq&C_m\alpha(k^{-1}+\alpha)^{m}
\\\label{ind5}\|(1+x)^{-m+1}\left(
\phi^{m}(A+B,j,\mathbf{k},\cdot)-\Phi_m\right)\|_{\infty}&\leq&C_m\alpha(k^{-1}+\alpha)^{m-1}
 \;.
\end{eqnarray}
For $m=0$ these equations hold (remember that
$f^0=\chi(j,\mathbf{k},\cdot)$) due to Theorem \ref{theo}.

Next we show, that $M-1$ implies $M$. Assume, that
(\ref{ind1})-(\ref{ind5}) hold for all $m< M$. Let us verify first
(\ref{ind1}) for $M$. For (\ref{deff}) we can write
\begin{eqnarray*}
f^{M}(A+B,j,\mathbf{k},\cdot)&:=&\partial_k^M \chi(j,\mathbf{k},\cdot)- \sum_{l=2}^M\left(%
\begin{array}{c}
  M \\
  j \\
\end{array}%
\right)
\partial_k^lT^{A+B}_{E_k}\partial_k^{(M-l)}\phi(A+B,j,\mathbf{k},\cdot)\\&&+M\partial_kT^{A+B}_{E_k}\phi^{M-1}(A+B,j,\mathbf{k},\cdot)\;.
\end{eqnarray*}
For compactly supported $A+B$ one has in view of (\ref{kernel}) for
any $\chi\in L^\infty$ that
\begin{eqnarray}\label{partialknT}
\|(1+x)^{-M+1}\partial_k^M T^{A+B}_{E_k}\chi\|_\infty&\leq&
\|\chi\|_\infty\sup_{\mathbf{x}\in\mathbb{R}^3}\left|(1+x)^{-M+1}\int
|\partial_k^NG^{+}_{E}(\mathbf{x}-\mathbf{y})|(A(\mathbf{y})+B(\mathbf{y}))\right|
\nonumber\\&\leq& C\|\chi\|_\infty\;.
\end{eqnarray}
Hence using (\ref{ind4}) and (\ref{ind5}) for $m< M$ it follows that
\begin{eqnarray*}
&&\|(1+x)^{-M+1}f^{M}(A+B,j,\mathbf{k},\cdot)\|_\infty\\&&\;\;\;=
C+\sum_{l=2}^MC_l \alpha(k^{-1}+\alpha)^{M-l-1}
+M\|\partial_kT^{A+B}_{E_k}\left(\phi^{M-1}(A+B,j,\mathbf{k},\cdot)-\Phi_{M-1}\right)\|_\infty
\\&&\;\;\;\;\;\;+M\|\partial_kT^{A+B}_{E_k}\Phi_{M-1}\|\;.
\\&&\;\;\;\leq C+C\alpha(k^{-1}+\alpha)^{M-l-1}+M\|\partial_kT^{A+B}_{E_k}\Phi_{M-1}\|\;.
\end{eqnarray*}
Below we will show, that $[\partial_k
T^{A}_{E_k}\Phi]_{k=0}=\lambda(\Phi)$ (see (\ref{dktphi})), hence
\begin{eqnarray*}
\partial_k T^{A+B}_{E_k}\Phi&=&\partial_k T^{A}_{E_k}\Phi+\partial_k T^{B}_{E_k}\Phi=
[\partial_k
T^{A}_{E_k}\Phi]_{k=0}+\mathcal{O}(k)+\mathcal{O}(\|B\|_1)\\&=&\mathcal{O}(k)+\mathcal{O}(\|B\|_1)
\end{eqnarray*}
for all $\Phi\in\mathcal{N}$. Hence
\begin{eqnarray*}
\|(1+x)^{-M+1}
f^M(A+B,j,\mathbf{k},\cdot)\|_{\infty}&\leq&C+C\alpha(k^{-1}+\alpha)^{M-2}+C(k+\|B\|_1)\alpha(k^{-1}+\alpha)^{M-1}\;.
\end{eqnarray*}
Note that $\alpha>1$ and $(k^{-1}+\alpha)^{-1}<k$, hence
\begin{eqnarray*}
\|(1+x)^{-M+1}
f^M(A+B,j,\mathbf{k},\cdot)\|_{\infty}&\leq&C(k+\|B\|_1)\alpha(k^{-1}+\alpha)^{M-1}\;,
\end{eqnarray*}
which is (\ref{ind1}) for $m=M$. It follows that also
$\|g_{M+1}\|_\infty\leq C (k+\|B\|_1) \alpha(k^{-1}+\alpha)^{M-1}$.
With Corollary \ref{cor2a} we get in view of (\ref{gengeq}) that
\begin{eqnarray*}
\|\Phi_m\|&\leq&C_m\alpha^2(k^{-1}+\alpha)^{m-2}\leq C_m\alpha(k^{-1}+\alpha)^{m-1}
\end{eqnarray*}and\begin{eqnarray*}
\|(1+x)^{-m+1}\left( \phi^{m}(A+B,j,\mathbf{k},\cdot)-\Phi_m\right)\|_{\infty}&\leq&
C_m\alpha(k^{-1}+\alpha)^{m-1}
 \;.
\end{eqnarray*}
which are (\ref{ind4}) and (\ref{ind5}) for $m=M$.

Induction over $m$ yields, that (\ref{ind1}) - (\ref{ind5}) hold for
all $m\in\mathbb{N}_0$.

With (\ref{ind4}) and (\ref{ind5}) the Theorem follows easily. Since
$\alpha>1$ we have that (i) if $\alpha>k^{-1}$ the right hand sides
of (\ref{ind4}) and (\ref{ind5}) are bounded by $C_m
2^m\alpha^{m+1}$, (ii) if $\alpha\leq k^{-1}$ the right hand sides
of (\ref{ind4}) and (\ref{ind5}) are bounded by $C_m 2^m\alpha
k^{-m}$ and the Theorem follows.

$\square$

Again we get in a similar but easier way the respective Theorem for
$\overline{\lambda}=1$.

\begin{thm}\label{lambda1der}
Let  $A\in\mathcal{C}$ with $\overline{\lambda}=1$. Then the
respective statement of Theorem \ref{lambda0der} holds with
$$\alpha=1+\left(\inf_{\Phi\in
\mathcal{N},\|\Phi\|=1}\left|\left\langle
\Phi,B,\Phi\right\rangle\right|+ k
 \right)^{-1}\;.$$
\end{thm}

As above one can make it easier to understand the statement of
Theorem \ref{theo}, by restriction on potentials $B_\mu$ which can
be written as $B_\mu\mu B_0$ for some fixed potential $B_0$ and
$\mu\in[-\mu_0,\mu_0]$.
\begin{corollary}\label{supercorder}
Let $A\in\mathcal{C}$ with $\overline{\lambda}=0$. Let $B_0\in
L^\infty\cap L^1$ with $\left\langle\Phi, B_0,\Phi\right\rangle\neq
0$ for all $\Phi\in\mathcal{N}\backslash\{0\}$. Then there exist
constants
 $C,\mu_0,k_0>0$ and constants $\gamma_l$, $l=1,\ldots,m\leq\dim\mathcal{N}$
  such that for any $m\in\mathbb{N}_0$ there exist $C_m<\infty$ such
 that for any
$\mathbf{k}\in\mathbb{R}^3$ with $k<k_0$, $j=1,2$, any
$\mu\in[-\mu_0,\mu_0]$ there exists a
$\Phi^\mu_{j,\mathbf{k}}\in\mathcal{N}$ with
\begin{equation*}\|(1+x)^{-m}\partial_k^m\phi(A+\mu B_0,j,\mathbf{k},\cdot)\|_\infty\leq
C_m\left(k^{-m}+\left|\sum_{l=1}^n\frac{k}{|\mu+\gamma_lk^2|+k^3}\right|^{m+1}\right)\;.
\end{equation*}
\end{corollary}

This Corollary can now be used to give estimates on the behavior of
the generalized eigenfunctions for critical potentials multiplied
with a factor close to one (i.e. considering the case
$A+B=\lambda\mu A$ with $\mu\approx 1$). Such potentials are a
comparably easy model to estimate physical processes under the
influence of critical fields with small perturbations. Therefore the
literature on Adiabatic Pair Creation (see e.g.
\cite{nenciu1,nenciu2}) deals with potentials $A$ multiplied by a
switching factor. We shall give a result suitable for such
application, imposing further conditions on the potential $A$ which
allow us to extend the bounds on $\mathbf{k}\in\mathbb{R}^3$. The
following Corollary shall play an important role in the proof of
adiabatic pair creation which has been achieved recently
\cite{pdneu}
\begin{corollary}
Let $A\in\mathcal{C}$ be positive and purely electric with
$\overline{\lambda}=0$. Then there exist constants
 $C,\delta>0$ and constants $\gamma_l$, $l=1,\ldots,m\leq\dim\mathcal{N}$
  such that for any $m\in\mathbb{N}_0$ there exist $C_m<\infty$ such
 that for any
$\mathbf{k}\in\mathbb{R}^3$, $j=1,2$, any
$\mu\in[1-\delta,1+\delta]$ there exists a
$\Phi^\mu_{j,\mathbf{k}}\in\mathcal{N}$ with
\begin{equation}\label{endcor}\|(1+x)^{-m}\partial_k^m\phi(\mu A,j,\mathbf{k},\cdot)\|_\infty\leq
C_m\left(k^{-m}+\left|\sum_{l=1}^n\frac{k}{|\mu+\gamma_lk^2|+k^3}\right|^{m+1}\right)\;.
\end{equation}
Furthermore there exist $\Phi_\mu(\mathbf{k},j,\cdot)\in\mathcal{N}$
and  $C$ uniform in $\mathbf{k}\in\mathbb{R}^3$ and
$\mu\in[1-\delta,1+\delta]$ so that
\begin{equation}\label{resultat2e}
\|\phi(\mu A,j,\mathbf{k},\cdot)-\Phi_\mu(\mathbf{k},j,\cdot)\|_\infty<C\;.
\end{equation}
\end{corollary}

\noindent\textbf{Proof:} For $k$ smaller than $k_0$ the Corollary
follows from Corollary \ref{supercorder} and Corollary
\ref{supercor} replacing $\mu$ by $1-\mu$ and setting $B_0=A$. Note,
that for positive $A$ one has $\left\langle \Phi, A, \Phi
\right\rangle>0$ for all $\Phi\in\mathcal{N}$, hence the assumptions
on $B_0$ in Lemma \ref{supercorder} are satisfied for $A$.

Using continuity of the operator $T$ one can find a uniform bound on
$\|\phi(\mu A,j,\mathbf{k},\cdot)\|_\infty$ for $\mathbf{k}$ in an
arbitrary compact subset of $\mathbb{R}^3$ not containing $k=0$ (see
for example \cite{pickl}). In \cite{pickl} it is also proven that
the left hand side of (\ref{endcor}) is bounded for
$k\rightarrow\infty$ and the Corollary follows.

$\square$

\section*{Acknowledgments} The author would like to thank Detlef D\"urr for suggesting to work on the topic. The discussions with him were very
helpful and influenced the paper a lot.

 Financial support by the
Austrian Science Fund in the form of an Erwin Schr\"odinger
Fellowship is gratefully acknowledged.

\section*{Appendix}

\subsection*{Control of $\Phi_1$ in (\ref{vorwirdnull1})}

Above we showed, that any element $\Phi\in \mathcal{N}$ is in $L^2$
if and only if $\lambda(\Phi)=0$. There we split
$\Phi=\Phi_1+x^{-1}\lambda(\Phi)$. Assuming that
$\Phi_1(\mathbf{x})$ decays at least as fast as $x^{-2}$ and using
$\Phi\in\mathcal{B}\subset L^\infty$ it in fact follows that
$\Phi\in L^2\Leftrightarrow \lambda(\Phi)=0$.

Let us now proof that under our assumptions of the potential we
always have that $\Phi_1(\mathbf{x})$ decays at least as fast as
$x^{-2}$.

For $\Phi_{1}(\mathbf{x})$ (c.f (\ref{vorwirdnull1})) we can write
defining $f(\mathbf{x}):=(1+x)^{2}A(\mathbf{x})
 \Phi(\mathbf{x})$
\begin{eqnarray*}
|\Phi_1(\mathbf{x})|
%
&\leq&\left|\int_{y<\frac{x}{2}}
\frac{1}{4\pi}\left(\frac{y-x}{yx}(1+\beta)+i\sum_{j=1}^{3}\alpha_{j}\frac{y_{j}}{y^3})\right)A(\mathbf{x}-\mathbf{y})
 \Phi(\mathbf{x}-\mathbf{y})d^{3}y\right|
\nonumber\\&&+\left|\int_{y\geq\frac{x}{2}}
\frac{1}{4\pi}\left(\frac{y-x}{yx}(1+\beta)
+i\sum_{j=1}^{3}\alpha_{j}\frac{y_{j}}{y^3})\right)A(\mathbf{x}-\mathbf{y})
 \Phi(\mathbf{x}-\mathbf{y})d^{3}y\right|
\\&\leq&\left|\int_{y<\frac{x}{2}}
\frac{1}{4\pi}(1+|\mathbf{x}-\mathbf{y}|)^{-2}\left(\frac{y-x}{yx}(1+\beta)+i\sum_{j=1}^{3}\alpha_{j}\frac{y_{j}}{y^3})\right)f(\mathbf{x}-\mathbf{y})d^{3}y\right|
\nonumber\\&&+\left|\int_{y\geq\frac{x}{2}}
\frac{1}{4\pi}\left(\frac{1}{yx}(1+\beta)
+i\sum_{j=1}^{3}\alpha_{j}\frac{y_{j}}{y^3})\right)f(\mathbf{x}-\mathbf{y})d^{3}y\right|
\\&\leq&\left(1+\frac{x}{2}\right)^{-2}\left|\int_{y\leq1}
\frac{1}{4\pi}\left(\frac{y-x}{yx}(1+\beta)+i\sum_{j=1}^{3}\alpha_{j}\frac{y_{j}}{y^3})\right)f(\mathbf{x}-\mathbf{y})d^{3}y\right|
\nonumber\\&&+\left(1+\frac{x}{2}\right)^{-2}\left|\int_{1<y<\frac{x}{2}}
\frac{1}{4\pi}\left(\frac{y-x}{yx}(1+\beta)+i\sum_{j=1}^{3}\alpha_{j}\frac{y_{j}}{y^3})\right)f(\mathbf{x}-\mathbf{y})d^{3}y\right|
\nonumber\\&&+\frac{2}{x^2}\left|\int_{y\geq\frac{x}{2}}
\frac{1}{4\pi}f(\mathbf{x}-\mathbf{y})d^3y\right|
\\&\leq&\left(1+\frac{x}{2}\right)^{-2}\|f\|_\infty\left|\int_{y\leq1}
\frac{1}{4\pi}\left(\frac{y-x}{yx}(1+\beta)+i\sum_{j=1}^{3}\alpha_{j}\frac{y_{j}}{y^3})\right)d^{3}y\right|
\nonumber\\&&+\left(1+\frac{x}{2}\right)^{-2}\left|\int_{1<y<\frac{x}{2}}
\frac{1}{4\pi}\left(3+\frac{2}{x}\right)f(\mathbf{x}-\mathbf{y})d^{3}y\right|
+\frac{2}{x^2}\left|\int_{y\geq\frac{x}{2}}
\frac{1}{4\pi}f(\mathbf{x}-\mathbf{y})d^3y\right|\;.
\end{eqnarray*}
Since $(1+x)^2A\in L^1\cap L^\infty$ and $\Phi\in\mathcal{B}\subset
L^\infty$ we have that $f\in  L^1\cap L^\infty$. Thus
$\Phi_1(\mathbf{x})$ decays like $x^{-2}$, it follows that
$\Phi_1\in L^2$.

\subsection*{Proof of Lemma \ref{haupt}}


Next we shall prove the following Lemma, the last points of which
are exactly Lemma \ref{haupt}.

\begin{lemma}\label{hauptalles}

Let $A\in\mathcal{C}$. Then there exist constants
 $C,C'>0$, $C_0,C_1\in\mathbb{R}$, a selfadjoint sesquilinear map $r:\mathcal{N}\times\mathcal{N}\rightarrow\mathbb{C}$
 and an anti-selfadjoint sesquilinear map
 $s:\mathcal{N}\times\mathcal{N}\rightarrow\mathbb{C}$ with $s(\chi,\chi)\neq0$ for all $\chi\in\mathcal{N}$ such that for any $\mathbf{k}\in\mathbb{R}^3$ with
$k<1$, any potential $B$ with $B\in L^1\cap L^\infty$ and any
normalized $h\in L^\infty$, normalized $m^\bot\in\mathcal{M}^\bot$
and normalized $\Phi,\Psi\in\mathcal{N}$

\begin{itemize}

\item[(a)]

\begin{equation*}
\| \left( T^{A}_{E_k}-1\right)m^\bot\|_\infty\geq C \;,
\end{equation*}%

\item[(b)]

\begin{equation*}
\|(T^{A}_{E_k}-T^A_1)h\|_\infty<Ck\;,
\end{equation*}

\item[(c)]

\begin{equation*}
\| P^\bot_\mathcal{M}h\|_\infty\leq C\;,
\end{equation*}

\item[(d)]

\begin{eqnarray*}
\|(T^A_{E_k}-T^A_1)\Phi\|_\infty&\leq& C(\overline{\lambda}  k+k^2)
\;,
\end{eqnarray*}

\item[(e)]

\begin{equation*}
\mid \left\langle
\Phi,A,(T^{A+B}_{E_k}-T^{A+B}_1)h\right\rangle\mid<C(\|A\|_1+\|B\|_1)(\overline{\lambda}
k+k^2)\;,
\end{equation*}

\item[(f)]
$$|\left\langle \Phi,A,(1-T^{A+B}_{E_k})m^\bot\right\rangle|<C (\|A\|_1+\|B\|_1)(\overline{\lambda}  k+k^2)\;,$$

\item[(g)]

$$\| P^\bot_\mathcal{M} (1-T^{A+B}_{E_k})m^\bot\|_\infty\geq C-C'(\overline{\lambda}  k+k^2+\|B\|_1+\|B\|_\infty)\;,$$

\item[(h)]

$$\| P^\bot_\mathcal{M} (1-T^{A+B}_{E_k})\Phi\|_\infty< C (\overline{\lambda} k+k^2+\|B\|_1+\|B\|_\infty)\;,$$

\item[(i)]

\begin{equation}\label{haupthformel1}
\left\langle \Phi,A, (T^{B}_1-T^{B}_{E_k})\Psi\right\rangle\leq\|
B\|_1\left(\overline{\lambda}
\mathcal{O}(k)+\mathcal{O}(k^2)\right)\end{equation}

and if $B=A$

\begin{equation}\label{haupthformel}\left\langle \Phi,A, (T^{A}_1-T^{A}_{E_k})\Psi\right\rangle=- ik \langle
\Phi,A,\lambda(\Psi) \rangle+r(\Phi,\Psi)
k^2+s(\Phi,\Psi)k^3+\text{o}(k^3)\;,\end{equation}

\item[(j)]

\begin{eqnarray*}\left\langle \Phi,A, (1-T^{A+B}_{E_k})\Psi\right\rangle
&=& \left\langle \Phi,B,\Psi\right\rangle- ik \langle
\Phi,A,\lambda(\Psi) \rangle+r(\Phi,\Psi)
k^2+s(\Phi,\Psi)k^3\\&&+\text{o}(k^3)+\|B\|_1( \overline{\lambda}
\mathcal{O}(k)+\mathcal{O}(k^2))\;.\end{eqnarray*}

\end{itemize}

\end{lemma}

\vspace{0.5cm} \noindent\textbf{Proof of (a)} Let $k\in\mathbb{R}$,
$m^\bot\in\mathcal{M}^\bot$. We will prove part (a) of the Lemma by
contradiction. Assume that for every $n\in\mathbb{N}$ there exists a
$k_0\leq k_n< 1$
 and a function $h_n\in\mathcal{M}^\bot$ with $\|h_n\|_\infty=1$ such that
\begin{equation}\label{asumeinv2}
\| \left(1- T^{A}_{E_{k_n}}\right)h_n\|_\infty<\frac{1}{n}\;,
\end{equation}
i.e.
\begin{equation*}
\lim_{n\rightarrow\infty}\left(1-T^{A}_{E_{k_n}}\right)h_n=0\;.
\end{equation*}
Using Bolzano Weierstra\ss\space we can assume without loss of
generality that $k_n$ converges. We denote the respective limit by
$k_0$. Using that $T^{A}_{E_k}$ is completely continuous it follows
that
\begin{equation}\label{limhn}
\lim_{n\rightarrow\infty}\left(1-T^{A}_{E_{k_0}}\right)h_n=0\;.
\end{equation}
But the sequence $T^{A}_{E_{k_0}} h_n$ is Arzela-Ascoli compact,
since
\begin{equation}\label{defseta}
\mathcal{A}:=\{T_{E_{k_0}}^{A} g\text{ with
}g\in\mathcal{B},\space\| g\|_\infty=1\}
\end{equation}
 is compact in the Arzela-Ascoli
sense, i.e.  for any $\varepsilon>0$ there exists a $\delta>0$ such
that
\begin{equation}\label{arzasc}
\mid f(\mathbf{x})-f(\mathbf{y})\mid <\varepsilon
\end{equation}
for all $\mathbf{x}, \mathbf{y}\in\mathbb{R}^3$ with $\|
\mathbf{x}-\mathbf{y}\|<\delta$ and all $f\in\mathcal{A}$.

To prove this let $\varepsilon>0$, $f\in\mathcal{A}$ and let
$k\in\mathbb{R}$ and $g\in\mathcal{B}$ be such that
$f=T_{E_{k_0}}^{A} g$ ,$\| g\|_\infty=1$.

Then
\begin{eqnarray}
\mid f(\mathbf{x})-f(\mathbf{y})\mid&=& \mid
T_{E_{k_0}}^{A}g(\mathbf{x})-T_{E_{k_0}}^{A}g(\mathbf{y})\mid\nonumber\\&=&\big|\int
G^{+}_{E_{k_0}}(\mathbf{x}-\mathbf{z})A(\mathbf{z})g(\mathbf{z})d^3z
-\int
G^{+}_{E_{k_0}}(\mathbf{y}-\mathbf{z})A(\mathbf{z})g(\mathbf{z})d^3z\big|
\nonumber\\&=&\left|\int
\left(G^{+}_{E_{k_0}}(\mathbf{x}-\mathbf{z})-G^{+}_{E_{k_0}}(\mathbf{y}-\mathbf{z})\right)A(\mathbf{z})g(\mathbf{z})d^3z
\right|\;.
\end{eqnarray}
 For any $\zeta>0$ we can write
\begin{eqnarray*}
\mid f(\mathbf{x})-f(\mathbf{y})\mid&\leq&\left|\int_{z<\zeta}
\left(G^{+}_{E_{k_0}}(\mathbf{x}-\mathbf{z})-G^{+}_{E_{k_0}}(\mathbf{y}-\mathbf{z})\right)A(\mathbf{z})g(\mathbf{z})d^3z
\right|
\\&&+\left|\int_{z>\zeta}
\left(G^{+}_{E_{k_0}}(\mathbf{x}-\mathbf{z})-G^{+}_{E_{k_0}}(\mathbf{y}-\mathbf{z})\right)A(\mathbf{z})g(\mathbf{z})d^3z
\right|
\nonumber\\&\leq&\| A(\mathbf{z})\|_\infty \|
g(\mathbf{z})\|_\infty\left|\int_{z<\zeta}
\left(G^{+}_{E_{k_0}}(\mathbf{x}-\mathbf{z})-G^{+}_{E_{k_0}}(\mathbf{y}-\mathbf{z})\right)d^3z
\right|
\\&&+
\sup_{r>\zeta}G^{+}_{E_{k_0}}(\mathbf{x}-\mathbf{r})-G^{+}_{E_{k_0}}(\mathbf{y}-\mathbf{r})\|
A(\mathbf{z})\|_1\| g\|_\infty\;.
\end{eqnarray*}
Since $G^{+}_{E_{k_0}}$ is integrable, the first summand goes to
zero in the limit $\zeta\rightarrow 0$. Hence we can find a
$\zeta>0$ such that the first summand is smaller than
$\varepsilon/2$.

Since $G^{+}_{E_{k_0}}$ is on any set bounded away from $0$
uniformly continuous, the second summand goes for any fixed
$\zeta>0$ to zero in the limit $|\mathbf{x}-\mathbf{y}|\rightarrow
0$. Hence we can find for any $\zeta>0$ a $\delta>0$ such that the
second summand is smaller than $\varepsilon/2$. It follows that
$\mid f(\mathbf{x})-f(\mathbf{y})\mid<\varepsilon$ for $\|
\mathbf{x}-\mathbf{y}\|<\delta$.

It follows that $\mathcal{A}$ is compact (in the Arzela-Ascoli
sense).

Thus there exists a convergent subsequence $$ (T_{E_{k_{n(j)}}}^{A}
h_{n(j)})_{j\in\mathbb{N}}$$ of $(T_{E_{k_0}}^{A}
h_n)_{n\in\mathbb{N}}$ with $\lim_{j\rightarrow\infty}
T_{E_{k_{n(j)}}}^{A} h_{n(j)}=h\in\mathcal{A}$ (i.e.
$\|h\|_\infty=1$).

By virtue of (\ref{limhn})
$\lim_{j\rightarrow\infty}h_{n(j)}=\lim_{j\rightarrow\infty}T_{E_{k_0}}^{A}h_{n(j)}=h$
and $(1-T_{E_{k_0}}^{A})h=0$. Since $(1- T_{E_{k_0}}^{A})h=0$ has
nontrivial solutions only for $k_0=0$ it follows that $k_0=0$ and
$h\in\mathcal{N}$.

On the other hand since $h_n\in\mathcal{M}^\bot$
\begin{equation*}
\left\langle \mathcal{N},A, h_n \right\rangle=0
\end{equation*}
for all $n\in\mathbb{N}$. With the continuity of the scalar product
it follows that $h\in\mathcal{M}^\bot$, which contradicts to the
fact that $\mathcal{N}\cap\mathcal{M}^\bot=\{0\}$ and part a) of the
Lemma follows.

\vspace{0.5cm} \noindent\textbf{Proof of (b)} Let $h\in L^\infty$,
$A\in \mathcal{C}$.  We have using (\ref{deft})
\begin{eqnarray*}
(T^A_{E_k}-T^A_1)h&=&\int
 \left(G^{+}_{E_k}(\mathbf{y})-G^{+}_1(\mathbf{y})\right)A(\mathbf{x}-\mathbf{y})h(\mathbf{x}-\mathbf{y})d^3y\;.
\end{eqnarray*}
It follows that
\begin{eqnarray*}
\|(T^A_{E_k}-T^A_1)h\|_\infty&\leq&\| h\|_\infty\int
 \mid G^{+}_{E_k}(\mathbf{y})-G^{+}_1(\mathbf{y})\mid
 A(\mathbf{x}-\mathbf{y})d^3y\\
 &\leq&\| h\|_\infty\| \frac{|\cdot|+1}{|\cdot|}A\|_1
\|
\frac{|\cdot|}{|\cdot|+1}(G^{+}_{E_k}(\cdot)-G^{+}_1(\cdot))\|_\infty
\;.
\end{eqnarray*}
Note that since $A\in L^1\cap L^\infty$ the $\|
\frac{|\cdot|+1}{|\cdot|}A\|_1$ exists.

Using the definition of $G^{+}_{E_k}$ (see (\ref{kernel})) we have
that
\begin{eqnarray*}&&\|\frac{|\cdot|}{|\cdot|+1} (G^{+}_{E_k}(\cdot)-G^{+}_1(\cdot))\|_\infty
\\&&\hspace{0.5cm}=\|\frac{1}{4\pi}\frac{e^{ikx}}{x+1}\left(-(E_{k}+\sum_{j=1}^{3}\alpha_{j}k\frac{x_{j}}{x}+\beta )
-ix^{-1}\sum_{j=1}^{3}\alpha_{j}\frac{x_{j}}{x}\right)\\&&\hspace{0.7cm}-\frac{1}{4\pi}\frac{1}{x+1}\left(-(1+\beta)
-ix^{-1}\sum_{j=1}^{3}\alpha_{j}\frac{x_{j}}{x}\right)\|_\infty
\\&&\hspace{0.5cm}\leq\|\frac{1}{4\pi}\frac{e^{ikx}}{x+1}\left(-(E_{k}-1+\sum_{j=1}^{3}\alpha_{j}k\frac{x_{j}}{x})\right)\|_\infty
\\&&\hspace{0.7cm}+\|
\frac{1}{4\pi}\frac{e^{ikx}-1}{x+1}\left(-(1+\beta)
-ix^{-2}\sum_{j=1}^{3}\alpha_{j}\frac{x_{j}}{x}\right)\|_\infty\;.
\end{eqnarray*}
The first summand is of order $k$. Since $e^{ikx}-1$ is of order
$kx$, the second summand is of order $k$ and part (b) of the Lemma
follows.

\vspace{0.5cm} \noindent\textbf{Proof of (c)} The triangle
inequality yields
\begin{eqnarray*}
\| P^\bot_\mathcal{M}h\|_\infty&\leq&\|
P^\parallel_\mathcal{M}h\|_\infty+\| h\|_\infty\;.
\end{eqnarray*}
Since $\mathcal{M}^\parallel$ has finite dimension, all norms on
this space are equivalent, i.e. there exists a $C<0$ such that
\begin{eqnarray*}\| P^\parallel_\mathcal{M}h\|_\infty&\leq& C\|
P^\parallel_\mathcal{M}h\|
=:C\sup_{\Phi\in\mathcal{N}}\|A\Phi\|^{-1}\int
A(\mathbf{x})\Phi(\mathbf{x})h(\mathbf{x})d^3x
\\&\leq&C\|h\|_\infty \|A\|_1\sup_{\Phi\in\mathcal{N}}\|A\Phi\|^{-1}\|
\Phi\|_\infty\;.
\end{eqnarray*}
Using the equivalence of all norms on the finitely dimensional
vector-space $\mathcal{M}^\parallel$ we have that $\|A\Phi\|^{-1}\|
\Phi\|_\infty$ is bounded and part (c) of the Lemma follows.

\vspace{0.5cm} \noindent\textbf{Proof of (i)} Let
$\Phi\in\mathcal{N}$ with $\| \Phi\|_\infty=1$. Using linearity it
suffices to prove equation (\ref{haupthformel1}) for $B$ with
$\|B\|_1=1$.

We shall use Taylors formula to estimate $\left\langle
\Phi,B,(T^{A}_1-T^{a}_{E_k})\Psi\right\rangle$. In view of
(\ref{deft})
\begin{equation*}T^A_{E_{k}}\Phi=\int G^{+}_{E_k}(\mathbf{y})A(\mathbf{x}-\mathbf{y})\Phi(\mathbf{x}-\mathbf{y})d^3y\;,
\end{equation*}
i.e. we develop $G^{+}_{E_k}$ (see (\ref{kernel})) around $k=0$, so
we need the following derivatives
\begin{eqnarray}\label{difG}
\partial_kG^{+}_{E_k}&=&\partial_k\left(\frac{1}{4\pi}e^{ikx}\left(-x^{-1}(E_{k}+\sum_{j=1}^{3}\alpha_{j}k\frac{x_{j}}{x}+\beta ) -ix^{-2}\sum_{j=1}^{3}\alpha_{j}\frac{x_{j}}{x}\right)\right)
\nonumber\\&=&\frac{e^{ikx}}{4\pi}\left(-i(E_{k}+\sum_{j=1}^{3}\alpha_{j}k\frac{x_{j}}{x}+\beta
)
+x^{-1}\sum_{j=1}^{3}\alpha_{j}\frac{x_{j}}{x}-x^{-1}(\frac{k}{E_k}+\sum_{j=1}^{3}\alpha_{j}\frac{x_{j}}{x})\right)
\nonumber\\&=&\frac{e^{ikx}}{4\pi}\left(-i(E_{k}+\sum_{j=1}^{3}\alpha_{j}k\frac{x_{j}}{x}+\beta
) -x^{-1}\frac{k}{E_k}\right)\;,
\end{eqnarray}
\begin{eqnarray}\label{difG2}
\partial_k^2G^{+}_{E_k}&=&\partial_k\left(\frac{1}{4\pi}e^{ikx}\left(-i(E_{k}+\sum_{j=1}^{3}\alpha_{j}k\frac{x_{j}}{x}+\beta )
-x^{-1}\frac{k}{E_k}\right)\right)
\nonumber\\&=&\frac{e^{ikx}}{4\pi}\left(x(E_{k}+\sum_{j=1}^{3}\alpha_{j}k\frac{x_{j}}{x}+\beta
)
-i\frac{k}{E_k}-i\frac{k}{E_k}-i\sum_{j=1}^{3}\alpha_{j}\frac{x_{j}}{x}-x^{-1}\frac{1}{E_k^3}\right)
\nonumber\\&=&\frac{e^{ikx}}{4\pi}\left(x(E_{k}+\sum_{j=1}^{3}\alpha_{j}k\frac{x_{j}}{x}+\beta
)-2i\frac{k}{E_k}
-i\sum_{j=1}^{3}\alpha_{j}\frac{x_{j}}{x}-x^{-1}\frac{1}{E_k^3}\right)
\end{eqnarray}
and
\begin{eqnarray}\label{difG3}
\partial_k^3G^{+}_{E_k}&=&\partial_k\left(\frac{1}{4\pi}e^{ikx}\left(x(E_{k}+\sum_{j=1}^{3}\alpha_{j}k\frac{x_{j}}{x}+\beta )-2i\frac{k}{E_k}
-i\sum_{j=1}^{3}\alpha_{j}\frac{x_{j}}{x}-x^{-1}\frac{1}{E_k^3}\right)\right)
\nonumber\\&=&\frac{e^{ikx}}{4\pi}\big(ix^2(E_{k}+\sum_{j=1}^{3}\alpha_{j}k\frac{x_{j}}{x}+\beta
)+2x\frac{k}{E_k}
+\sum_{j=1}^{3}\alpha_{j}x_{j}-i\frac{1}{E_k^3}\nonumber\\&&+x\frac{k}{E_k}+
\sum_{j=1}^{3}\alpha_{j}x_{j}-2i\frac{1}{E_k^3}+3x^{-1}\frac{k}{E_k^5}\big)
\nonumber\\&=&\frac{e^{ikx}}{4\pi}\big(ix^2(E_{k}+\sum_{j=1}^{3}\alpha_{j}k\frac{x_{j}}{x}+\beta
)+3x\frac{k}{E_k}
\nonumber\\&&+2\sum_{j=1}^{3}\alpha_{j}x_{j}-3i\frac{1}{E_k^3}+3x^{-1}\frac{k}{E_k^5}\big)\;.
\end{eqnarray}
By Taylors formula we have that
\begin{eqnarray}\label{taylorsp2a}
\left\langle \Phi,B, T^{A}_{E_k}\Psi\right\rangle&=&
k\left[\partial_k\left\langle
\Phi,B,T^A_{E_k}\Psi\right\rangle\right]_{k=0}
+\frac{1}{2}k^2\left[\partial_k^2\left\langle
\Phi,B,T^A_{E_k}\Psi\right\rangle\right]_{k=0}+o(k^2)
\nonumber\\&=:&S_1+S_2+\text{o}(k^2)\;.
\end{eqnarray}
For $S_1$ we obtain with (\ref{difG}) that
\begin{eqnarray}\label{dktphi}
 [\partial_kT^A_{E_k}]_{k=0}\Psi&=&-i\int
 \frac{1}{4\pi}\left(1+\beta\right)A(\mathbf{x}-\mathbf{y})
 \Psi(\mathbf{x}-\mathbf{y})d^3y=\lambda(\Psi)\;.
\end{eqnarray}
Hence by (\ref{wirdnull})
\begin{equation}\label{vorsecond} S_1=-1k \left\langle \Phi,B,\lambda(\Psi)\right\rangle \;.
\end{equation}
For $S_2$ we have
\begin{eqnarray*}
S_2&=&\frac{1}{2}k^2\left\langle\Psi,B,\int
 [\partial_k^2
 G^{+}_{E_k}(\mathbf{x}-\mathbf{y})]_{k=0}A(\mathbf{y})\Phi(\mathbf{y})d^3y
\right\rangle\;.
\end{eqnarray*}
In view of (\ref{difG2}) we have that for any $k_0>0$ there exists a
$C>0$ such that
$$|[\partial_k^2G^{+}_{E_k}(\mathbf{x}-\mathbf{y})]_{k=0}|\leq C (|\mathbf{x}-\mathbf{y}|+|\mathbf{x}-\mathbf{y}|^{-1})$$
uniform in $k<k_0$. Hence
\begin{eqnarray*}
\left|\int
 \partial_k^2
 G^{+}_{E_k}(\mathbf{x}-\mathbf{y})A(\mathbf{y})\Phi(\mathbf{y})d^3y\right|
&\leq&\int_{|\mathbf{x}-\mathbf{y}|<1}
2C|\mathbf{x}-\mathbf{y}|^{-1}|A(\mathbf{y})\Phi(\mathbf{y})|d^3y
\\&&+\int_{|\mathbf{x}-\mathbf{y}|>1}
 2C|\mathbf{x}-\mathbf{y}||A(\mathbf{y})\Phi(\mathbf{y})|d^3y
\\&\leq&C\|A\|_\infty\|\Phi\|_\infty+C\int_{|\mathbf{x}-\mathbf{y}|>1}
 (x+y)|A(\mathbf{y})\Phi(\mathbf{y})|d^3y\;.
\end{eqnarray*}
Since $(1+|\cdot|)A\in L^1$ and $\Phi\in L^\infty$ it follows that
there exists a $C>0$ such that
\begin{eqnarray*}
\int
 \partial_k^2
 G^{+}_{E_k}(\mathbf{x}-\mathbf{y})A(\mathbf{y})\Phi(\mathbf{y})d^3y&\leq&C(1+x)\;.
 \end{eqnarray*}
 Hence
\begin{eqnarray*}
|S_2|&\leq&\frac{1}{2}k^2C\left\langle\Psi,B,(1+x)\right\rangle\;.
\end{eqnarray*}
Using that $B\in L^1$ and that $(1+x)\Psi\in L^\infty$ (see below
(\ref{wirdnull})) (\ref{haupthformel1}) follows.

Next we prove (\ref{haupthformel}). We have by Taylors formula that
\begin{eqnarray}\label{taylorsp2}
\left\langle \Phi,A, T^{A}_{E_k}\Psi\right\rangle&=&
k\left[\partial_k\left\langle
\Phi,A,T^A_{E_k}\Psi\right\rangle\right]_{k=0}
\nonumber\\&&+\frac{1}{2}k^2\left[\partial_k^2\left\langle
\Phi,A,T^A_{E_k}\Psi\right\rangle\right]_{k=0}
\nonumber\\&&+\frac{1}{6}k^3\left[\partial_k^3\left\langle
\Phi,A,T^A_{E_k}\Psi\right\rangle\right]_{k=0}
\nonumber\\&&+\text{o}(k^3)
\nonumber\\&=:&k\widetilde{q}(\Phi,\Psi)+k^2r(\Phi,\Psi)+k^3s(\Phi,\Psi)+\text{o}(k^3)\;.
\end{eqnarray}
Setting $B=A$ in the estimates above (see (\ref{vorsecond}) and
below) we get that there exists a $C_1\in\mathbb{R}$ such that
$\widetilde{q}(\Phi,\Psi)=- ik \langle \Phi,A,\lambda(\Psi) \rangle$
and that $k^2r$ is well defined. Similarly we can show that $s$ is
well defined, now using that $(1+x^2)A\in L^1$.

Using the symmetry of the operator $T_{E_k}^{A}$ and the symmetry of
$i\partial_k$ we have that $r$ is selfadjoint and $s$ is
anti-selfadjoint.

It is left to show, that $s(\chi,\chi)\neq0$ for all
$\chi\in\mathcal{N}$. Let $\chi\in\mathcal{N}$. We obtain by
(\ref{difG3})
\begin{eqnarray}\label{splits3}
s(\chi,\chi)&=&-\frac{1}{6} \left\langle
\int
 \partial_k^3
 G^{+}_{E_k}(\mathbf{x}-\mathbf{y})A(\mathbf{y})\chi(\mathbf{y})d^3y
,A,\chi\right\rangle
\nonumber\\&=&-\frac{1}{6} \left\langle \frac{1}{4\pi}
\int
i(\mathbf{x}-\mathbf{y})^2(1+\beta )
 A(\mathbf{y})\chi(\mathbf{y})d^3y
,A,\chi\right\rangle
\nonumber\\&&-\frac{1}{6} \left\langle \frac{1}{4\pi}
\int
2\sum_{j=1}^{3}\alpha_{j}(x_{j}-y_j)
 A(\mathbf{y})\chi(\mathbf{y})d^3y
,A,\chi\right\rangle
\nonumber\\&&-\frac{1}{6} \left\langle \frac{1}{4\pi}
\int
3i\frac{1}{m^3}
 A(\mathbf{y})\chi(\mathbf{y})d^3y
,A,\chi\right\rangle
%
%
\nonumber\\&=&-\frac{ i}{24\pi} \int\int
A(\mathbf{x})(\mathbf{x}-\mathbf{y})^2
 A(\mathbf{y})\chi^\dagger(\mathbf{y})(1+\beta)\chi(\mathbf{x})d^3y
d^3x
\nonumber\\&&-\frac{1}{12\pi} \int\int A(\mathbf{x})\sum_{j=1}^{3}
 A(\mathbf{y})\chi^\dagger(\mathbf{y})\alpha_{j}(x_{j}-y_j)\chi(\mathbf{x})d^3y
d^3x
\nonumber\\&&+\frac{  i}{8\pi m} \int\int A(\mathbf{x})
 A(\mathbf{y})\chi^\dagger(\mathbf{y})\chi(\mathbf{x})d^3y
d^3x
\nonumber\\&=:&s_1+s_2+s_3\;.
\end{eqnarray}
For $s_1$ we can write
\begin{eqnarray}\label{s5split}
s_1&=&-\frac{ i}{24\pi} \int\int
A(\mathbf{x})(\mathbf{x}^2+\mathbf{y}^2)
 A(\mathbf{y})\chi^\dagger(\mathbf{y})(1+\beta)\chi(\mathbf{x})d^3y
d^3x
\nonumber\\&&+\frac{i}{12\pi} \int\int
A(\mathbf{x})\mathbf{x}\cdot\mathbf{y}
 A(\mathbf{y})\chi^\dagger(\mathbf{y})(1+\beta)\chi(\mathbf{x})d^3y
d^3x\;.
%
%
\end{eqnarray}
Using symmetry in exchanging $\mathbf{x}$ with $\mathbf{y}$ on the
first term it becomes
\begin{eqnarray*}
&&-\frac{i}{12\pi} \int\int A(\mathbf{x})\mathbf{x}^2
 A(\mathbf{y})\chi^\dagger(\mathbf{y})(1+\beta)\chi(\mathbf{x})d^3y
d^3x
\\&=&-\frac{i}{12\pi} \int A(\mathbf{x})\mathbf{x}^2\chi^\dagger(\mathbf{x})\int(1+\beta)
 A(\mathbf{y})\chi(\mathbf{y})d^3y
d^3x=0
\end{eqnarray*}
by (\ref{wirdnull}). Thus
\begin{eqnarray*}
s_1&=&\frac{i}{12\pi} \int
A(\mathbf{x})\chi^\dagger(\mathbf{x})\mathbf{x}d^3x(1+\beta)
\cdot\int\mathbf{y}
 A(\mathbf{y})\chi(\mathbf{y})d^3y\;.
\end{eqnarray*}
Setting
\begin{equation}\label{neustern}
\xi:=(12\pi)^{-1/2}\int A(\mathbf{x})\chi(\mathbf{x})\mathbf{x}d^3x
\end{equation}
we obtain
\begin{eqnarray}\label{neustern2}
s_1&=&i  \left\langle\xi (1-\beta), \xi\right\rangle\;.
\end{eqnarray}
Since $\beta$ is self adjoint it follows that
$\left\langle\xi(1-\beta), \xi\right\rangle\in\mathbb{R}$, since $\|
\beta \|=1$ it follows that $\left\langle\xi (1-\beta),
\xi\right\rangle\geq0$ hence there exists a $C_2\in\mathbb{R}_0^+$
such that
\begin{equation}\label{s5b}
s_1=i C_2\;.
\end{equation}
Due to symmetry in exchanging $\mathbf{x}$ with $\mathbf{y}$ we have
that
\begin{equation}\label{s6}
s_2=-s_2=0\;.
\end{equation}
For $s_3$ we can write
\begin{eqnarray}\label{s4gr01}
s_3&=&\frac{  i}{8\pi} \left|\int
A(\mathbf{x})\chi(\mathbf{x})d^3x\right|^2\;,
\end{eqnarray}
it follows that there exists a $C_3\geq0$ with
\begin{equation}\label{s4gr0}
s_3=i C_3\;.
\end{equation}
This (\ref{s5b}) and (\ref{s6}) in (\ref{splits3}) yield that there
exists a $C_1\geq0$ such that
\begin{equation}\label{fourthsumm}
s(\chi,\chi)=i C_1\;.
\end{equation}
Since $A$ was defined to satisfy either (\ref{restricta1}) or
(\ref{restricta2}) it follows taking note of (\ref{neustern}) and
(\ref{neustern2}) as well as (\ref{s4gr01}) that $C_2$ or $C_3>0$,
hence $C_1=C_2+C_3>0$, i.e. $s(\chi,\chi)\neq0$.

\vspace{0.5cm}
\noindent\textbf{Proof of part (d) of Lemma \ref{hauptalles}}\\

Similar as above we have using Taylors formula that
\begin{eqnarray*}
(T^A_{E_k}-1)\Phi&=& (T^A_1-1)\Phi+k[\partial_k(  T^A_{E_k})\Phi]_{k=0}+\mathcal{O}(k^2)
\;.
\end{eqnarray*}
Since $\Phi\in\mathcal{N}$
$$(T^A_1-1)\Phi=0\;.$$
It follows that
\begin{eqnarray*}
(  T^A_{E_k}-1)\Phi&=&k[\partial_k(
T^A_{E_k}-1)\Phi]_{k=0}+\mathcal{O}(k^2)\| \Phi\|_\infty\nonumber
\end{eqnarray*}
and
\begin{eqnarray}\label{taylor2}
\|(  T^A_{E_k}-1)\Phi\|_\infty&\leq&k\|[\partial_k(
T^A_{E_k}-1)\Phi]_{k=0}\|_\infty+\mathcal{O}(k^2)\| \Phi\|_\infty\;.
\end{eqnarray}
With (\ref{difG}) and (\ref{deft}) we have that by virtue of
(\ref{wirdnull})
\begin{eqnarray*}
[\partial_k(  T^A_{E_k}-1)]_{k=0}\Phi&=&\frac{-i}{4\pi}\int
 (1+\beta)A(\mathbf{y})\Phi(\mathbf{y})d^3y\;.
\end{eqnarray*}
Using (\ref{wirdnull}) it follows that
\begin{eqnarray*}
[\partial_k(
T^A_{E_k}-1)\Phi]_{k=0}&=&\frac{i\lambda(\Phi)}{4\pi}\;.
\end{eqnarray*}
With  (\ref{taylor2}) part (d) follows.

\vspace{0.5cm} \noindent\textbf{Proof of (e)} Using (\ref{symm}) and
part (d) of the Lemma yields
\begin{eqnarray*}
\mid \left\langle
(T^{A+B}_{E_k}-T^{A+B}_1)h,A,\Phi\right\rangle\mid&=&\mid
\left\langle h,A+B,(T^A_{E_k}-T^A_1)\Phi\right\rangle\mid
\\&\leq&\| (A+B)h \|_1 \|(T^A_{E_k}-T^A_1)\Phi\|_\infty
\\&\leq&\|h\|_\infty \|A+B\|_1 C (\overline{\lambda} k+k^2)\;.
\end{eqnarray*}
Using the triangle inequality part (e) of the Lemma follows.

\vspace{0.5cm} \noindent\textbf{Proof of (f)} Using (\ref{symm}) we
have
\begin{eqnarray*}
\left\langle \Phi,A,(1-T^{A+B}_{E_k})m^\bot\right\rangle
&=&\left\langle \Phi,A,(1-T^{A+B}_1)m^\bot\right\rangle+\left\langle
\Phi,A,(T^{A+B}_1-T^{A+B}_{E_k})m^\bot\right\rangle
\\&=&\left\langle (1-T^{A}_1)\Phi,A+B,m^\bot\right\rangle+\left\langle \Phi,A,(T^{A+B}_1-T^{A+B}_{E_k})m^\bot\right\rangle
\\&=&\left\langle \Phi,A,(T^{A+B}_1-T^{A+B}_{E_k})m^\bot\right\rangle\;.
\end{eqnarray*}
In view of part (e) we get part (f) of the Lemma.

\vspace{0.5cm} \noindent\textbf{Proof of (g)} Using the triangle
inequality and linearity of $P^\bot_\mathcal{M}$ and $T^{A+B}_{E_k}$
and (\ref{symm}) we have that
\begin{eqnarray}\label{partfzer}
\| P^\bot_\mathcal{M} (1-T^{A+B}_{E_k})m^\bot\|_\infty&\geq&
\|  (1-T^{A+B}_{E_k})m^\bot\|_\infty
-\| P^\parallel_\mathcal{M} (1-T^{A+B}_{E_k})m^\bot\|_\infty
\nonumber\\&\geq&\| (1-T^{A}_{E_k})m^\bot\|_\infty\nonumber
-\| T^{B}_{E_k}m^\bot\|_\infty
\\&&-\| P^\parallel_\mathcal{M} (1-T^{A}_{E_k})m^\bot\|_\infty
\nonumber\\&=:&S_1-S_2-S_3\;.
\end{eqnarray}
Using part (a) of the Lemma we have that
\begin{equation}\label{s1grc}
S_1\geq C\;.
\end{equation}
For $S_2$ we have
\begin{eqnarray*}
S_2&=&\| T^{B}_{E_k}m^\bot\|_\infty=\|\int
 G^{+}_{E_k}(\mathbf{x}-\mathbf{y}) B(\mathbf{y})
 m^\bot(\mathbf{y})d^{3}y\|_\infty
 \\&\leq&\|\int_{|\mathbf{x}-\mathbf{y}|<1}
 G^{+}_{E_k}(\mathbf{x}-\mathbf{y}) B(\mathbf{y})
 m^\bot(\mathbf{y})d^{3}y\|_\infty
 \\&&+\|\int_{|\mathbf{x}-\mathbf{y}|>1}
 G^{+}_{E_k}(\mathbf{x}-\mathbf{y}) B(\mathbf{y})
 m^\bot(\mathbf{y})d^{3}y\|_\infty\;.
\end{eqnarray*}
Since $G^+(\mathbf{x})$ is integrable for all $k<k_0$ and bounded
uniform in $k<k_0$ and $x>1$ it follows that there exists a constant
$C$ such that
\begin{equation}\label{s2blaa}
S_2\leq C\|B\|_\infty+C\|B\|_1\;.
\end{equation}
For $S_3$ we use part (f) of the Lemma. Choose $\Phi$ in part (f)
such that $A\Phi$ is parallel to
$P^\parallel_\mathcal{M}(1-T^{A}_{E_k})m^\bot$ and normalized. It
follows that
\begin{eqnarray*}
|\left\langle
\Phi,A,(1-T^{A+B}_{E_k})m^\bot\right\rangle|=|\left\langle
\Phi,A,P^\parallel_\mathcal{M}(1-T^{A+B}_{E_k})m^\bot\right\rangle+\left\langle
\Phi,A,P^\bot_\mathcal{M}(1-T^{A+B}_{E_k})m^\bot\right\rangle|\;.
\end{eqnarray*}
Using the definition of $P^\bot_\mathcal{M}$ the second summand is
zero, hence (remember that $\Phi$ was defined such that $A\Phi$ is
parallel to $P^\parallel_\mathcal{M}(1-T^{A+B}_{E_k})m^\bot$)
\begin{eqnarray*}
|\left\langle
\Phi,A,(1-T^{A+B}_{E_k})m^\bot\right\rangle|=|\left\langle
\Phi,A,P^\parallel_\mathcal{M}(1-T^{A+B}_{E_k})m^\bot\right\rangle|=\|
P^\parallel_\mathcal{M}(1-T^{A+B}_{E_k})m^\bot\| =S_3\;.
\end{eqnarray*}
Using the equivalence of all norms on the finite dimensional
vector-space $\mathcal{M}^\parallel$ and part (f) of the Lemma it
follows that there exists a $C>0$ such that $S_3<C
(\|A\|_1+\|B\|_1)(\overline{\lambda}  k+k^2)$. With
(\ref{partfzer}), (\ref{s1grc}) and (\ref{s2blaa}) part (g) of the
Lemma follows.

\vspace{0.5cm} \noindent\textbf{Proof of part (h)} Since
$\Phi\in\mathcal{N}$, i.e. $\Phi=T^A_1\Phi$ it follows with part (c)
of the Lemma that there exists a $C>0$ such that
\begin{eqnarray}\label{partfzer2}
\| P^\bot_\mathcal{M} (1-T^{A+B}_{E_k})\Phi\|_\infty&=&
\| P^\bot_\mathcal{M} (T^A_1-T^{A+B}_{E_k})\Phi\|_\infty
\nonumber\\&\leq&C\|(T^A_1-T^{A+B}_{E_k})\Phi\|_\infty
\nonumber\\&\leq&C\|(T^A_1-T^{A}_{E_k})\Phi\|_\infty+C\|(T^A_{E_k}-T^{A+B}_{E_k})\Phi\|_\infty
\nonumber\\&=:&S_1+S_2\;.
\end{eqnarray}
For $S_1$ we have using part (d) of the Lemma that there exists a
$C>0$ such that
\begin{equation}\label{s1grc2}
S_1\leq C(\overline{\lambda} k+k^2)\;.
\end{equation}
$S_2$ can be estimated similarly as $S_2$ above. We have
\begin{eqnarray*}
S_2&=&C\|(T^A_{E_k}-T^{A+B}_{E_k})\Phi\|_\infty=C\|
T^{B}_{E_k}\Phi\|_\infty
\\&=&C\|\int
 G^{+}_{E_k}(\mathbf{x}-\mathbf{y}) B(\mathbf{y})
 \Phi(\mathbf{y})d^{3}y\|_\infty
 \\&\leq&C\|\int_{|\mathbf{x}-\mathbf{y}|<1}
 G^{+}_{E_k}(\mathbf{x}-\mathbf{y}) B(\mathbf{y})
 \Phi(\mathbf{y})d^{3}y\|_\infty
 \\&&+C\|\int_{|\mathbf{x}-\mathbf{y}|>1}
 G^{+}_{E_k}(\mathbf{x}-\mathbf{y}) B(\mathbf{y})
 \Phi(\mathbf{y})d^{3}y\|_\infty\;.
\end{eqnarray*}
Since $G^+(\mathbf{x})$ is integrable for all $k<k_0$ and bounded
uniform in $k<k_0$ and $x>1$ it follows that there exists a constant
$C$ such that
\begin{equation*}
S_2\leq C\|B\|_\infty+C\|B\|_1\;.
\end{equation*}
With (\ref{partfzer2}) and (\ref{s1grc2})  part (h) of the Lemma
follows.

\vspace{0.5cm} \noindent\textbf{Proof of (j)} Using that
$\Psi\in\mathcal{N}$, i.e. $\Psi=T_1^A\Psi$ and linearity of
$T^A_{E_k}$ in $A$ we get
\begin{eqnarray}\label{final}
\left\langle \Phi,A, (1-T^{A+B}_{E_k})\Psi\right\rangle&=&\nonumber
\left\langle \Phi,A, (T^A_1-T^{A+B}_{E_k})\Psi\right\rangle
\\\nonumber&=&\left\langle \Phi,A, (T^{A+B}_1-T^{A+B}_{E_k})\Psi\right\rangle
+\left\langle \Phi,A, (T^A_1-T^{A+B}_1)\Psi\right\rangle
\\\nonumber&=&\left\langle \Phi,A, (T^{A}_1-T^{A}_{E_k})\Psi\right\rangle
+\left\langle \Phi,A, (T^{B}_1-T^{B}_{E_k})\Psi\right\rangle
-\left\langle \Phi,A, T^{B}_1\Psi\right\rangle\;.
\end{eqnarray}
Note, that due to (\ref{symm})
$$\left\langle \Phi,A, T^{B}_1\Psi\right\rangle=\left\langle \Phi,B, T^{A}_1\Psi\right\rangle=\left\langle \Phi,B,
\Psi\right\rangle\;.$$ Using this and (\ref{haupthformel1}) on the
first, (\ref{haupthformel}) on the second summand in (\ref{final})
(remember, that we need results for fixed $A$ and rather general
$B$, hence the $\|(1-x)^2 A\|$ dependence is in the constants)
yields part (j) of the Lemma.

\end{document}